\documentclass[12pt,onecolumn,draftclsnofoot]{IEEEtran}

\usepackage{subfig}
\usepackage{flushend}
\input{makrosFoundations}
\ifx\SetFigFont\undefined%
\makeatletter
\newcommand\SetFigFont[5]{%
  \reset@font\fontsize{#1}{#2pt}%
  \fontfamily{#3}\fontseries{#4}\fontshape{#5}%
  \selectfont}%
\makeatother
\fi

\newcommand\inputsinrsubfloat[1]{
\begin{picture}(0,0)%
\includegraphics{#1}%
\end{picture}%
\setlength{\unitlength}{3947sp}%
\begin{picture}(3600,2100)(0,-12370)
\put(3354,-12170){\makebox(0,0)[lb]{\smash{{\SetFigFont{12}{14.4}{\rmdefault}{\mddefault}{\updefault}{\color[rgb]{0,0,0}$n$}%
}}}}
\put(-50,-11336){\rotatebox{90.0}{\makebox(0,0)[lb]{\smash{{\SetFigFont{12}{14.4}{\rmdefault}{\mddefault}{\updefault}{\color[rgb]{0,0,0}SIR}}%
}}}}
\end{picture}%
}

\newcommand\inputtxpowersubfloat[1]{
\begin{picture}(0,0)%
\includegraphics{#1}%
\end{picture}%
\setlength{\unitlength}{3947sp}%
\begin{picture}(3600,2100)(0,-12370)
\put(3530,-12170){\makebox(0,0)[lb]{\smash{{\SetFigFont{12}{14.4}{\rmdefault}{\mddefault}{\updefault}{\color[rgb]{0,0,0}$n$}%
}}}}
\put(-70,-11636){\rotatebox{90.0}{\makebox(0,0)[lb]{\smash{{\SetFigFont{12}{14.4}{\rmdefault}{\mddefault}{\updefault}{\color[rgb]{0,0,0}transmit power}}%
}}}}
\end{picture}%
}
\graphicspath{{graphics/}}

\newcommand{\switchModes}[2]{#1}

\title{Decentralized Admission Control for Power-Controlled Wireless
  Links
}

\author{S\l awomir Sta\'nczak, Micha\l{} Kaliszan and Nicholas Bambos
  \thanks{S. Sta\'nczak and M. Kaliszan are with Fraunhofer German-Sino Lab
    for Mobile Communications Einsteinufer 37, D-10587 Berlin, Germany Email:
    \{slawomir.stanczak, michal.kaliszan\}@hhi.fraunhofer.de}\thanks{N. Bambos is
    with the Department of Electrical Engineering 350 Serra Mall, Stanford
    University, Stanford, CA 94305 Email: bambos@stanford.edu}\thanks{This work
    was presented in part at the 42nd Asilomar Conference on Signals, Systems, and
    Computers, Monterey, CA, USA.}}

\begin{document}
\maketitle

\begin{abstract}
  This paper deals with the problem of admission control/channel access in
  power-controlled decentralized wireless networks, in which the
  quality-of-service (QoS) is expressed in terms of the signal-to-interference
  ratio (SIR). We analyze a previously proposed admission control algorithm,
  which was designed to maintain the SIR of operational (active) links above some
  given threshold at all times (protection of active links). This protection
  property ensures that as new users attempt to join the network, the already
  established links sustain their quality. The considered scheme may be thus 
  applicable in some cognitive radio networks, where the fundamental premise
  is that secondary users may be granted channel access only if it does not
  cause disturbance to primary users.

  The admission control algorithm was previously analyzed under the assumption
  of affine interference functions. This paper extends all the previous
  results to arbitrary standard interference functions, which capture many
  important receiver designs, including optimal linear reception in the sense
  of maximizing the SIR and the worst-case receiver design.  Furthermore, we
  provide novel conditions for protection of active users under the considered
  control scheme when individual power constraints are imposed on each
  link. Finally, we consider the possibility of a joint optimization of
  transmitters and receivers in networks with linear transceivers, which
  includes linear beamforming in multiple antenna systems. Transmitter
  optimization is performed alternately with receiver optimization to generate
  non-decreasing sequences of SIRs. Numerical evaluations show that additional
  transmitter side optimization has potential for significant performance
  gains.
\end{abstract}



\section{Introduction}
\label{sec:intro}

Admission control is an important element of wireless communication systems
and its significance stems from the nature of most applications, where
arrivals of new users as well as departures of existing ones occur on an
irregular basis. Reallocating the available resources from the scratch after
such arrivals or departures is at best highly inefficient, in many cases
simply infeasible. This necessitates the development of admission control
schemes to decide whether or not a new user is allowed to join the network,
and how should the admission be organized so that the existing (active) users are
not disturbed. Recently, admission control has acquired additional importance
in the context of cognitive radio, where the necessary condition for allowing
secondary (unlicensed) users to access some given resources is that the operation 
or primary (licensed) users remains unhindered.

An interesting contribution in the field of admission control is the work
presented in \cite{Bambos00}, where the authors addressed the admission
control problem\footnote{In this paper, it may be more accurate to think of
  ``channel access'' rather than ``admission control''. We use the latter term
  as both concepts are closely related in decentralized networks.} in
power-controlled decentralized wireless networks. The main idea is the
introduction of an active link protection (ALP) mechanism to sustain the SIR
of active links above required thresholds, as new links attempt to access the
same channel by gradually increasing their transmit powers (in a guarded
manner).  The active links are endowed with an SIR protection margin to
cushion the effect of increased interference.

%


The work of \cite{Bambos00} was preceded by extensive research in the area of
power control. In particular, References \cite{Foschini93,Yates95a} proposed a
distributed asynchronous on-line power control algorithm, which can satisfy
user-specific SIR requirements at the minimum transmitter powers. It converges
geometrically fast to a global unique optimum, provided that the users have
feasible SIR requirements. Yates \cite{Yates95a} further showed that the
convergence result remains valid if the interference power at each receiver
output is any standard (interference) function of the transmit powers. The
class of standard interference functions is described by three
(non-restrictive) axioms, namely positivity, scalability and
monotonicity. Such an interference model captures most practical receiver
designs, including the worst-case receiver design and optimal linear reception
in the sense of maximizing the SIR. It is also general enough to incorporate
cross-layer effects, and it serves as a theoretical basis for many algorithms.
The analysis in \cite{Bambos00} considered linear interference, which may also
be viewed as a special case of interference functions introduced in
\cite{Yates95a}.

\subsection{Paper contribution and structure}
\label{sec:intro_contr}

In this paper, we extend the work of \cite{Bambos00} in the following three
directions:
\begin{description}
\item[(a)] We incorporate the axiomatic interference model of \cite{Yates95a} by
  assuming that the interference perceived by any user/link is characterized
  by a standard interference functions. This model includes linear
  interference functions assumed in \cite{Bambos00} as a special case.
\end{description}

It was already pointed out by \cite[Sect. V.C]{Yates95a} that the ALP property
of the algorithm in \cite{Bambos00}, which is the ability to maintain the SIR
values of the active links above a given SIR targets at all times, carries
over to standard interference functions.  In this paper, we show however that
the algorithm preserves its \emph{all} properties under standard interference
functions. All the results presented in \cite{Bambos00} can be thus derived
from the simple axiomatic framework. These results are presented in Sect.
\ref{sec:ActiveLinkProtection_ACL}.

\begin{description}
\item[(b)] We take into consideration individual power constraints on each link and
  prove sufficient conditions for the algorithm to provide the ALP.
\end{description}

As mentioned in \cite[Sect. V.C]{Yates95a} and \cite[Sect. VII]{Bambos00}, the
ALP is not preserved when the limitations on transmit powers are taken into
account. As a remedy, the authors of \cite{Bambos00} suggested equipping the
admission control scheme with a forced drop-out mechanism which causes
new/inactive links to drop out when they push active ones beyond their maximum
powers. In this paper, we provide novel conditions for having the ALP
property. Of particular interest may be ``on-line'' conditions that guarantee
the ALP, provided that the network is in a some state, which is shown to be
achieved in a finite time; as long as such a state is not achieved, no new
users are allowed to access the channel. These results can be found in Sect.
\ref{sec:RemarkPowerConstraints}.

\begin{description}
\item[(c)] We assume a wireless network equipped with optimal linear receivers in
  the sense of maximizing each SIR, which includes multiple antenna systems
  with optimal linear receive beamforming \cite{SchubBoche06NoW}. We
  investigate the impact of \emph{transmitter} side optimization (in addition
  to the receiver side optimization) when the links are not fully admissible.
\end{description}

Based on the ideas of \cite{UlukusWireless04}, transmitter optimization is
performed alternately with receiver optimization so as to generate a sequence
of non-decreasing SIRs. Numerical evaluations show that additional transmitter
side optimization has a potential for huge performance gains. The problem of
linear transceiver optimization is addressed in Sect.
\ref{sec:ActiveLinkProtection_Beamforming}.


\subsection{Further related work}
\label{sec:intro_RelatedWork}

The problem of power control in wireless networks has been an active research
area for more than the past two decades. Early works focused on centralized
and distributed power control, including the so-called max-min SIR balancing
problem (SIR: signal-to-interference ratio) and the QoS-based power control
aiming at satisfying given desired SIR levels (SIR targets) with a minimum
total transmit power. Both approaches have been extensively studied and are
fairly well understood
\cite{aein73,Zander92,Foschini93,Yates95a,Yates95b,gerlach96,Bambos98,Wu00,zander01,Elbatt04,ChiangBook2008PowerControl}.
The work evolved towards distributed power control algorithms (see for
instance \cite{Foschini93,Yates95a} and \cite{Hanly95} for combined power
control and cell-site selection). 
The axiomatic framework in \cite{Yates95a} allowed for extracting
more general properties of various power control algorithms. A slightly different
framework of general interference functions was proposed in \cite{SchubBoche06NoW}
and was used as a basis for analysis of properties of selected classes of 
interference functions \cite{BocheSchubert08ConcaveAndConvex,BocheSchubert08ACalculus}. 

The early works on power control focused on the convergence behavior of 
the algorithms. The DPC/ALP algorithm proposed in \cite{Bambos00} and analyzed
in this paper was one the first attempts to provide certain 
performance guarantees during the transient phase, before the algorithm converged. 
In \cite{Fazel06TransientAnalysis}, the authors use tools from control theory to analyze 
the behavior of the iterative
power control algorithm with linear interference (\cite{Foschini93}). This allows 
them to prove existence of invariant sets in the SIR domain and thus provide
conditions under which links are protected from dropping below the required SIR target.

A framework for adapting the transmission power under varying system
conditions so that the perceived QoS is guaranteed is considered in
\cite{Cuomo03improvingwireless}.  The authors of \cite{Elbatt04} present a
cross-layer design framework for contention-based wireless networks with
scheduling and power control phases combined in a alternating way. Reference
\cite{Alpcan08PowerControl} considered power control in multi-cell networks as
a team optimization problem and also analyzed admission control for the
presented model.  A different approach is followed in
\cite{Naqvi06ADistributedChannel}, where channel reservation is combined with
power control to eliminate the need for incremental power-up phase of the
inactive users. Reference \cite{Baccelli04UpAndDownlink} proposed admission
and congestion control scheme for large networks with homogeneous user
distribution, which takes both intra- and inter-cell interference into account
as well as power constraints. Admission control for the uplink of a multi-cell
wireless network is considered in \cite{JuiTengWang09AdmissionControl}. Two
approaches were proposed based on admission criteria or connection removal.
Other interesting and related contributions on the subject include
\cite{Marantz05AdmissionControl}, \cite{Liu04calladmission} and
\cite{ChoongMingChin06CallAdmission}.

The authors of \cite{Jurgens05AHomotopy} approached the admission control
problem for a network with both multiple user classes and multiple service
classes by formulating it as a homotopy method. In
\cite{Mathar08ProportionalQoS}, the authors suggested proportional reduction
of SIR requirements as a control mechanism in the case of overload.  Several
attempts have also been made to use Game Theory for the considered problem.
In \cite{Huang04auction-basedspectrum}, the authors presented auction-based
mechanisms for allocating power, which are capable of achieving a weighted
max-min fair SINR allocation or maximizing the total utility. Reference
\cite{HuangJSAC06} considered a distributed power control scheme in which each
user announces a price to be paid by other users for the interference they
cause.  The authors of \cite{Xiao01DistributedAdmission} analyzed the case of
linear interference functions and introduced a system parameter called
discriminant.  Two distributed protocols are considered, in which the value of
the discriminant is used to decide whether a new user can be admitted to the
system. References \cite{koskie05} and \cite{XiaoShroff03} proposed
distributed power control schemes in which some cost for each mobile is used.
Finally, in a recent publication \cite{TanPalChi07Exploiting}, the starting
point for the analysis is the algorithm proposed by \cite{Bambos00}, which is
also the starting point for our paper. The authors
\cite{TanPalChi07Exploiting} analyzed the tradeoff between energy consumption
and robustness.

The basic idea of
cognitive radio, proposed in \cite{MitolaThesis00}, is to identify highly
underutilized frequency bands of the radio spectrum and then allow
unlicensed users to access these bands. 
Various approaches to implementing cognitive radio
networks have been considered ever since. These approaches include among others
spectrum sensing and spectrum sharing using centralized or decentralized
cognitive MAC protocols; for an overview we refer to \cite{Akyildiz2006Next}.

There is a close connection between the concepts of cognitive radio and
admission control, as implementing a cognitive radio network in a distributed
environment requires an appropriate access scheme.
This approach is followed by the authors of \cite{Mitliagkas08ConvexApproximation}.
Primary users have a
certain QoS requirements that cannot be violated.  The goal is to maximize the
number of admitted secondary users, where each secondary user, if admitted,
also has a certain QoS requirement. The problem is NP-hard, and the authors
proposed a convex approximation algorithm.


\section{System model, assumptions and definitions}
\label{sec:ActiveLinkProtection_model}

We consider an arbitrary power controlled wireless network with $K$ (logical)
links that are referred to as \emph{users}. Let $\logic:=\{1,\dotsc,K\}$ and
let $\ve{p}=(p_1,\dotsc,p_K)\geq 0$ be the power vector (allocation), whose
$k$th coordinate is transmit power of user $k$ at some time instant. In this
paper, we assume that the wireless channel is arbitrary (chosen randomly) but
fixed. This is a reasonable assumption for a broad class of networks in which
channels vary slowly so that optimization algorithms need significantly less
time than the coherence time. The performance measure of interest is the
signal-to-interference ratio (SIR), which is defined to be
\begin{equation*}
  \sir_k(\ve{p})=p_k/I_k(\ve{p})\geq 0\,.
\end{equation*}
Here and hereafter, $I_k:\RN^K\to\RP$ is any standard interference function
that fulfills the following axioms:
\begin{definition}[Standard Interference Function \cite{Yates95a}]
\label{def:InterFunStandard}
We say that $I_k:\RN^K\to\R,k\in\logic,$ is a standard interference function
if each of the following holds.
 \begin{enumerate}[{$A$}1]
 \item $I_k(\ve{p}) >0$ for all $\ve{p}\geq 0$ (\emph{positivity}).
 \item $I_k(\mu\ve{p}) <\mu I_k(\ve{p})$ for any $\ve{p}\geq 0$ and $\mu>1$
   (\emph{scalability}).
 \item $I_k(\ve{p}^{(1)})\geq I_k(\ve{p}^{(2)})$ if $\ve{p}^{(1)}\geq \ve{p}^{(2)}$ 
 	(\emph{monotonicity}).
 \end{enumerate}
\end{definition}
It may be verified that the linear (affine) interference function assumed
in \cite{Foschini93,Bambos00} and given by
\begin{equation}
  \label{eq:LinearInterFun}
I_k(\ve{p})=(\vmat\ve{p}+\ve{z})_k=\sum\nolimits_{l\in\logic}v_{k,l}p_l+z_k 
\end{equation}
satisfies the axioms. Here, $\vmat=(v_{k,l})$ is the so-called gain matrix,
$v_{k,l}\geq 0$ with $v_{k,k}=0$ are (effective) power gains determined by the
transceiver structure, wireless fading channel etc, and
$\ve{z}=(z_1,\dotsc,z_K)>0$ is the noise vector.  Note that
$v_{k,l}=V_{k,l}/V_{k}\geq 0$, where $V_{k}>0$ is the signal power gain and
$V_{k,l}$ denotes the interference power gain, is independent of the power
allocation. This is for instance the case when each link employs the
matched-filter receiver or linear successive interference cancellation
receiver.

One can however go one step further in assuming that the power gains -- and
with it the interference powers -- depends on some \emph{adaptive} receive
strategy to be chosen depending on power allocation. Assuming the power vector
$\ve{p}\geq 0$ and some adjustable receive strategy\footnote{In general,
  $\menge{U}$ is a compact subset of the unit sphere chosen so as to take into
  account potential constraints on the receiver structure.}
$\ve{u}_k\in\menge{U}=\{\ve{u}:\|\ve{u}\|_2=1\}$, a more general model of the
(effective) interference power at the output of receiver $k$ is
\begin{equation}
  \label{eq:varrho}
  \varrho_k(\ve{p},\ve{u}_k)=\sum\nolimits_{l\in\logic}v_{k,l}(\ve{u}_k)p_l+z_k(\ve{u}_k)\,. 
\end{equation}
We see that the receive strategies influence the power gains $v_{k,l},l\neq
k,$ and the effective noise power $z_k$. It is pointed out that the $k$th
receive strategy impacts only the interference of user $k$. Now the
interference function under an optimal receiver in the sense of maximizing
each SIR for a given power vector is of the form
\begin{equation}
\label{eq:InterFunMMSE}
I_k(\ve{p})=\min\nolimits_{\ve{u}\in\menge{U}}
\varrho_k(\ve{p},\ve{u}),\;k\in\logic\,.
\end{equation}
It may be verified that this interference function satisfies the axioms of
Definition \ref{def:InterFunStandard} as well, and therefore is standard.

Other examples of standard interference functions are
$\RN^K\to\RN:\ve{p}\mapsto\max_{\xi\in\xset}\varrho_k(\ve{p},\xi)$ and
$\RN^K\to\RN:\ve{p}\mapsto\min_{u\in\menge{U}_k}\max_{\xi\in\xset}\varrho_k(\ve{p},u,\xi)$
where $\varrho_k(\ve{p},\xi)$ and $\varrho_k(\ve{p},u,\xi)$ are standard
interference functions of $\ve{p}$ for any fixed $\xi\in\xset$ and
$u\in\menge{U}_k$. Given a power allocation $\ve{p}$, the values
$\varrho_k(\ve{p},\xi)$ and $\varrho_k(\ve{p},u,\xi)$ are equal to the
interference powers under some interference uncertainty $\xi$ from some
(suitable) compact set $\xset$. The interference uncertainty means here that
the interference power continuously varies depending on the choice of
$\xi\in\xset$. Thus, interference functions of this form can be used, for
instance, to model the worst-case interference under imperfect channel
knowledge \cite{Jorswieck2004}.

Let $I(\ve{p})=(I_1(\ve{p}),\dotsc,I_K(\ve{p}))$. This vector-valued
interference function is referred to as standard if $I_k$ for each $k$ is a
standard interference function. The following proposition, whose proof is
omitted for lack of space, is an extension of the result presented in
\cite{SchubBoche06NoW} to nonnegative power vectors.
\begin{proposition}
\label{prop:InterFunContinuous}
Let $I:\RN^K\to\RP^K$ be a standard interference function. Then, $I$ is
component-wise continuous. 
\end{proposition}

Let $\gamma_k>0$ be the SIR target of user $k$ in the sense that this user is
satisfied with the quality-of-service provided by the network if
$\sir_k(\ve{p})\geq \gamma_k$ for some power vector $\ve{p}$. If this holds
for every user, then the power allocation is also said to be valid. Now,
considering Definition \ref{def:InterFunStandard} and Proposition
\ref{prop:InterFunContinuous}, it follows that
$\gmat=\diag(\gamma_1,\dotsc,\gamma_K)$ is feasible if and only if
\begin{equation}
\label{eq:FeasibilityCond}
0<C(\gmat):=\inf_{\ve{p}>0} \max_{k\in\logic}\frac{\gamma_k I_k(\ve{p})}{p_k}<1\,.
\end{equation}
Note that the infimum cannot be attained due to the axiom $A2$ but, by the
axiom $A1$, it must be larger than $0$.

By \cite{Yates95a}, we know that if $C(\gmat)<1$, then there exists a unique
power vector $\ve{p}^\ast>0$ such that
\begin{equation}
  \label{eq:fixedpoint_def}
  \ve{p}^\ast=\intfun(\ve{p}^\ast)
\end{equation}
where and hereafter (for brevity) we use 
$\intfun(\ve{p}):=(\intfunk_1(\ve{p}),\dotsc,\intfunk_K(\ve{p}))=(\gamma_1
I_1(\ve{p}),\dotsc,\gamma_K I_K(\ve{p}))$.  Moreover, from \cite{Yates95a}, we
know that the iteration
\begin{equation}
\label{eq:yatesIteration}
\ve{p}(n+1)=\intfun(\ve{p}(n)),\quad n\in\NNZ
\end{equation}
converges to $\ve{p}^\ast$ given by (\ref{eq:fixedpoint_def}) as $n\to\infty$,
regardless of the choice of $\ve{p}(0)$. In other words, if
(\ref{eq:FeasibilityCond}) is satisfied, then (\ref{eq:yatesIteration})
converges to the unique fixed point of the standard interference function
$\intfun$. Note that the SIR targets are satisfied with equality under the
power vector $\ve{p}^\ast>0$.

\section{Power Control with active link protection}
\label{sec:ActiveLinkProtection_ACL}

Let $n\in\NN$ be a time index. Given 
$\{\ve{p}(n)\}_{n\in\NN}=\{(p_1(n), \dotsc, p_K(n))\}_{n\in\NN}$,
a sequence of power vectors, we use
$\sir_k(n):=\sir_k(\ve{p}(n))$
to denote the SIR of user $k$ at time $n$. Let
$\emenge{A}_n=\{k\in\logic:\sir_k(n)\geq\gamma_k\}$ be the index set of users
that satisfy their SIR targets at time $n$.  Furthermore, we define
$\emenge{B}_n=\logic\setminus\emenge{A}_n$. We say that user $k$ is active at
time $n$ if $k\in\emenge{A}_n$. Otherwise, it is said to be inactive. Each
inactive user aims at becoming an active one. Without loss of generality, it
is assumed that $\emenge{A}_n=\{1,\dotsc,M_n\}$ and
$\emenge{B}_n=\{M_n+1,\dotsc,K\}$ for some $1\leq M_n\leq K$. Occasionally, we
need the following definition
$\bar{\emenge{B}}_n:=\emenge{B}_n-|\emenge{A}_n|=\{1,\dotsc,K-M_n\}$.

We consider the following admission control algorithm with active link
protection (ALP) for power-controlled networks \cite{Bambos00}:
\begin{equation}
  \label{eq:ActiveLinkProtection_ACL_Alg}
  p_k(n+1)=
  \begin{cases}
    \delta\,\gamma_k I_k(\ve{p}(n)) & k\in\emenge{A}_n\\
    \delta\,p_k(n)=\delta^{n+1}p_k(0) & k\in\emenge{B}_n
  \end{cases}
\end{equation}
where $\delta\in(1,\infty)$ is some given constant,
$\emenge{A}_0\neq\emptyset$ (at least one user is assumed to be admitted at
the beginning since otherwise we have a classical power control problem) and
$I_k$ is any standard interference function (according to Definition
\ref{def:InterFunStandard}) and $p_k(0)>0$ is arbitrary for each
$k\in\emenge{B}_0$.  Note that (\ref{eq:ActiveLinkProtection_ACL_Alg}) is
called an admission control algorithm as all users in $\emenge{B}_n$ are
seeking admission to the network. As $k\in\emenge{A}_n$ if and only if
$p_k(n)\geq\gamma_kI_k(\ve{p}(n))$, the iteration
(\ref{eq:ActiveLinkProtection_ACL_Alg}) can be equivalently written as
\begin{equation}
  \label{eq:AlgwithJn}
  \ve{p}(n+1)=\delta\aclfun(\ve{p}(n)) 
\end{equation}
where $\aclfun=(\aclfunk_{1},\dotsc,\aclfunk_{K}):\RN^{K}\to\RN^{K}$ is given
by
\begin{equation}
  \label{eq:Jn_Def}
  \begin{split}
    \aclfunk_{k}(\ve{p})=\min\bigl\{p_k,\gamma_k I_k(\ve{p})\bigr\}
    =\min\bigl\{p_k,\intfunk_k(\ve{p})\bigr\}\,.
  \end{split}
\end{equation}
We point out that unless $\emenge{B}_n=\emptyset$, $\aclfun$ is not a standard
interference function as the axioms $A1$ and $A2$ are not satisfied.  Thus,
the convergence of (\ref{eq:ActiveLinkProtection_ACL_Alg}) and
(\ref{eq:AlgwithJn}) does not follow from \cite{Yates95a} (note that in
\cite{Yates95a}, the corresponding interference functions are made standard by
adding an arbitrarily small positive vector to the power vector).

Throughout the paper, we use $\ve{p}^{(a)}(n)$ and $\ve{p}^{(i)}(n)$ to denote
the power vectors of the active and inactive users at time $n$,
respectively. Hence, $\ve{p}(n)=(\ve{p}^{(a)}(n),\ve{p}^{(i)}(n))$ and, if the
time index $n$ can be dropped, we have
$\ve{p}=(\ve{p}^{(a)},\ve{p}^{(i)})$. Moreover,
$\intfun^{(a)}_n:\RN^{K}\to\RN^{|\emenge{A}_n|}$ and
$\intfun^{(i)}_n:\RN^{K}\to\RN^{|\emenge{B}_n|}$ are used to
denote the corresponding interference functions. Consequently,
$\intfun_n(\ve{p})=(\intfun^{(a)}_n(\ve{p}),\intfun^{(i)}_n(\ve{p}))$ and
$\intfun(\ve{p})=(\intfun^{(a)}(\ve{p}),\intfun^{(i)}(\ve{p}))$. It is
important to notice that $\intfun_n\neq\aclfun$, unless
$\emenge{B}_n=\emptyset$ or, equivalently, unless all users are admitted to
the network. Following \cite{Bambos00}, we differentiate between the three
cases:
\begin{assumptions}
\item \label{as:fullyadmissible}
$C(\gmat)<C(\delta\gmat)<1$: The users seeking admission to the network
  (inactive users) are fully admissible.
\item \label{as:deltaincomp}
$C(\gmat)<1$ and $C(\delta\gmat)\geq 1$: The inactive users are fully
  admissible but $\delta$-incompatible.
\item \label{as:totallyinadmissible}
$C(\gmat)\geq 1$: The inactive users are not fully admissible or, using
  the terminology of \cite{Bambos00}, \emph{totally inadmissible}.
\end{assumptions}

\subsection{Some properties of the control scheme}
\label{sec:ActiveLinkProtection_results}

Now, we prove interesting properties of the scheme. Throughout this subsection,
we need the following lemma.
\begin{lemma}
\label{lem:DecreaseInterference}
Let $\delta>1$ be arbitrary. For any $n\in\NNZ$ and $k\in\logic$, we have
$I_k(\ve{p}(n+1))<\delta I_k(\ve{p}(n))$.
\end{lemma}
\begin{IEEEproof}
  Let $n\in\NNZ$ be arbitrary. As $p_{k}(n)\geq\gamma_{k} I_{k}(\ve{p}(n))$
  for any $k\in\emenge{A}_n$, it follows from the definition of $\aclfun$ in
  (\ref{eq:Jn_Def}) that $\aclfun(\ve{p}(n))\leq\ve{p}(n)$. So, by $A2$ and
  $A3$, this implies that
  \begin{align*}
    I_k(\ve{p}(n+1))&=I_k(\delta\aclfun(\ve{p}(n)))<\delta I_k(\aclfun(\ve{p}(n)))
    \switchModes{}{\\ &}\leq\delta I_k(\ve{p}(n)),\quad k\in\logic\,.
\end{align*}
The proof is complete. 
\end{IEEEproof}

Now we use the lemma to show an important property of the algorithm, namely
the \emph{protection of active users} or, in short, ALP. This property was
already reported in \cite[Theorem 10]{Yates95a}.
\begin{proposition}
\label{prop:ACL_ActiveStayActive}
Let $\delta>1$. Then, $\emenge{A}_n\subseteq\emenge{A}_{n+1},n\in\NNZ$.
\end{proposition}
\begin{IEEEproof}
  Let $n\in\NNZ$ and $k\in\emenge{A}_n$ be arbitrary. Since we have
  $p_k(n+1)=\delta\gamma_kI_k(\ve{p}(n))$, Lemma
  \ref{lem:DecreaseInterference} implies that
  \begin{align*}
    \sir_k(\ve{p}(n+1))&=\frac{p_k(n+1)}{I_k(\ve{p}(n+1))}
    >\frac{\delta\gamma_k I_k(\ve{p}(n))}{\delta I_k(\ve{p}(n))}=\gamma_k\,.
  \end{align*}
  Thus, $k\in\emenge{A}_{n+1}$, which completes the proof.
\end{IEEEproof}

Note that the proposition holds even if $C(\gmat)\geq 1$, that is, even if the
inactive users are totally inadmissible. In other words, the users seeking
admission to the network under the considered strategy do not destroy the
connections of the active users in the sense that the SIR targets of these
users remain satisfied. This protection is achieved at the cost of increased
transmit powers of the active users. The increase of power in every iteration
step is however bounded above by $\delta$. Indeed, for any $k\in\emenge{A}_n$
and $n\in\NN$, one has
\begin{equation}
  \label{eq:increasePowerActive_bounded}
  p_k(n+1)/p_k(n)<\delta\,.
\end{equation}
This is because if $k\in\emenge{A}_n$, then $p_k(n)\geq\gamma_k
I_k(\ve{p}(n))$, and hence $p_k(n+1)=\delta\gamma_k I_k(\ve{p}(n))\leq\delta
p_k(n)$. Moreover, by the proof of Proposition
\ref{prop:ACL_ActiveStayActive}, strict inequality holds yielding
(\ref{eq:increasePowerActive_bounded}). The next proposition shows that the
SIRs of inactive users increases under the power control iteration
(\ref{eq:ActiveLinkProtection_ACL_Alg}).
\begin{proposition}
\label{prop:ACL_SIRinactiveGrows}
Let $\delta>1$ and $k\in\emenge{B}_n\neq\emptyset$ be arbitrary. Then, for
every $n\in\NN$, $\sir_k(\ve{p}(n))<\sir_{k}(\ve{p}(n+1))$.
\end{proposition}
\begin{IEEEproof}
  Let $k\in\emenge{B}_n$ and $n\in\NNZ$ be arbitrary. Then, 
\begin{equation*}
  \sir_k(\ve{p}(n+1))=\frac{p_k(n+1)}{I_k(\ve{p}(n+1))}=\frac{\delta p_k(n)}{I_k(\ve{p}(n+1))}\,.
\end{equation*}
By Lemma \ref{lem:DecreaseInterference}, we have $I_k(\ve{p}(n+1))<\delta
I_k(\ve{p}(n))$. Consequently, $\sir_k(\ve{p}(n+1))>\sir_k(\ve{p}(n))$.
\end{IEEEproof}

Again, we point out that the proposition holds regardless of whether the
inactive users are fully admissible or not.  Note that the algorithm
generates a \emph{strictly} increasing sequence of SIRs for each user seeking
admission to the network. As a result, an inactive user either becomes an
active one or its SIR converges to some value $\bar{\sir}_k<\gamma_k$
(due to the boundedness and strict increasingness of the sequence).

In the remainder of this section, we assume the following additional
condition on interference functions.
\begin{assumptions}
\item \label{as:NonOrthogonal}
If $\emenge{B}_n\neq\emptyset$ for some $n\in\NNZ$, then, for each
  $k\in\emenge{A}_n$, there is $l\in\emenge{B}_n$ such that $I_k$ is strictly
  increasing in $p_l$. 
\end{assumptions}

The above condition means that no active user is orthogonal to all inactive
users. Thus, we exclude the trivial cases when the inactive users have no
impact on active ones.

\begin{lemma}
\label{lem:convGeneralFun}
For each $k\in\logic$, we have
$\lim_{c\to\infty} \intfunk_k(c\ve{p})/c=\gintfunk_k(\ve{p}),\ve{p}\geq 0$,
for some function $\gintfunk_k:\RN^K\to\RN$ satisfying each of the following:
 \begin{enumerate}[{$\tilde{A}$}1]
 \item $\gintfunk_k(\ve{p})\geq 0$ (nonnegativity).
 \item $\gintfunk_k(\mu\ve{p})=\mu \gintfunk_k(\ve{p})$ for all $\mu>0$
   (homogeneity).
 \item $\gintfunk_k(\ve{p}^{(1)})\geq \gintfunk_k(\ve{p}^{(2)})$
   if $\ve{p}^{(1)}\geq \ve{p}^{(2)}$ (monotonicity).
      \end{enumerate}   
\end{lemma}
\begin{IEEEproof}
The proof can be found in App. \ref{app:proofLemmaAsym}. 
\end{IEEEproof}

The Lemma \ref{lem:convGeneralFun} states that $\intfunk_k(c\ve{p})/c$
converges to a general interference function \cite{SchubBoche06NoW} as $c$
tends to infinity.  In particular, for the linear interference function
(\ref{eq:LinearInterFun}) and the minimum interference function
(\ref{eq:InterFunMMSE}) this is in accordance with the intuition that the 
additive background noise can be neglected if the power vector is scaled to
infinity. An important consequence of the lemma is the fact that the function
$\gintfunk_k(\ve{p})=\lim_{c\to\infty}\intfunk_k(c\ve{p})/c$ is continuous
for $\ve{p}>0$ \cite[Sect. 2.1.2]{SchubBoche06NoW}. Note that assuming
a strictly positive power vector $\ve{p}$ does not restrict generality as by
$A1$ and \eqref{eq:ActiveLinkProtection_ACL_Alg} we have $p_k(n)>0$ 
for all $k\in\logic$ and $n>0$.

\begin{lemma}
\label{lem:extendIntIsStandard}
If \ref{as:NonOrthogonal} holds,
$\gintfunk_k((\ve{p}^{(a)},\ve{p}^{(i)})),k\in\emenge{A}$, is a
standard interference function of $\ve{p}^{(a)}\geq 0$ for any fixed
$\ve{p}^{(i)}>0$.
\end{lemma}
\begin{IEEEproof}
  $A$3 of Definition \ref{def:InterFunStandard} follows directly from
  $\tilde{A}$3. Positivity $A$1 is due to \ref{as:NonOrthogonal} and the fact
  that $\ve{p}^{(i)}$ is positive. Indeed, if there was
  $\ve{p}=(\ve{p}^{(a)},\ve{p}^{(i)})\geq 0$ with $\ve{p}^{(i)}>0$ such that
  $\gintfunk_k(\ve{p})=0$, then, by \ref{as:NonOrthogonal}, we would
  obtain $\gintfunk_k(\mu\ve{p})<0$ for any $\mu\in(0,1)$, which would
  contradict $\tilde{A}1$. Scalability $A$2 is a consequence of $\tilde{A}$2
  and \ref{as:NonOrthogonal}: For any $\mu>1$, we have
  $\gintfunk_k((\mu\ve{p}^{(a)},\ve{p}^{(i)}))
  =\gintfunk_k(\mu(\ve{p}^{(a)},1/\mu\ve{p}^{(i)}))
  =\mu\gintfunk_k((\ve{p}^{(a)},1/\mu\ve{p}^{(i)}))
  <\mu\gintfunk_k((\ve{p}^{(a)},\ve{p}^{(i)}))$ where the last step is
  due to \ref{as:NonOrthogonal} and $\mu>1$.
\end{IEEEproof}

It is pointed out that Lemma \ref{lem:extendIntIsStandard} can be easily
deduced from \cite{BocheSchubertICC09AUnifying}. Now we use Lemmas
\ref{lem:convGeneralFun} and \ref{lem:extendIntIsStandard} to prove the
following result.

\begin{proposition}
  \label{prop:ACL_TotallyInadmissible}
  Suppose that \ref{as:totallyinadmissible} and 
  \ref{as:NonOrthogonal} hold. Let $\emenge{A}=\cup_{n\in\NNZ}\emenge{A}_n$,
  $\emenge{B}=\cap_{n\in\NNZ}\emenge{B}_n\neq\emptyset$, and
  \begin{align*}
  \gintfun^{(a)}(\ve{p}):=(\gintfunk_k(\ve{p}))_{k\in\emenge{A}}
  &&\text{with}&&
  \gintfunk_k(\ve{p})=\lim_{c\to\infty}\intfunk_k(c\ve{p})/c\,. 
  \end{align*}
  Then, as $n\to\infty$, $\sir_k(n)\to\bar{\sir}_k\in(0,\infty)$ and
  $p_k(n)/\delta^n\to\bar{p}_k\in(0,\infty)$. If $k\in\emenge{B}$, then
  $\bar{p}_k=p_k(0)>0$.  In contrast, for each $k\in\emenge{A}$, we have
  \begin{equation*}
    \bar{\sir}_k=\frac{\gamma_k\bar{p}_k}{\gintfunk_k(\bar{\ve{p}})}=
    \frac{\gamma_k\bar{p}_k}{\gintfunk_k((\bar{\ve{p}}^{(a)},\ve{p}^{(i)}(0)))}
    =\gamma_k, k\in\emenge{A}
\end{equation*}
where $\ve{p}^{(i)}(0)=(p_k(0))_{k\in\emenge{B}}>0$ and
$\bar{\ve{p}}^{(a)}=(\bar{p}_k)_{k\in\emenge{A}}$ satisfies
$\bar{\ve{p}}^{(a)}=\gintfun^{(a)}((\bar{\ve{p}}^{(a)},\ve{p}^{(i)}(0)))$.
\end{proposition}
\begin{IEEEproof}
  The proof is deferred to Appendix \ref{app:proofTotallyInAdmissible}. 
\end{IEEEproof}

Now we replace the condition of total inadmissibility
\ref{as:totallyinadmissible} by full admissibility \ref{as:fullyadmissible} to
show that the algorithm (\ref{eq:ActiveLinkProtection_ACL_Alg}) does what it
was designed to do.

\begin{proposition}
  \label{prop:ACL_Totallyadmissible}
  Suppose that \ref{as:fullyadmissible} and \ref{as:NonOrthogonal} hold. Then,
  there is a finite $n_0\in\NN$ so that $\emenge{A}_{n_0}=\logic$.  Moreover,
  as $n\to\infty$, we have $p_k(n)\to \bar{p}_k=\delta\gamma_k
  I_k(\bar{\ve{p}})=\delta\intfunk_k(\bar{\ve{p}}),k\in\logic$.
\end{proposition}
\begin{IEEEproof}
The reader can find the proof in Appendix \ref{app:proofTotallyAdmissible}. 
\end{IEEEproof}

The last result considers the case \ref{as:deltaincomp}. 

 \begin{proposition}
  \label{prop:ACL_NotDeltaCompatible}
  Let \ref{as:deltaincomp}, \ref{lem:convGeneralFun} and \ref{as:NonOrthogonal}
  be satisfied. Then, there is a finite $n_0\in\NN$ so that
  $\emenge{A}_n=\logic$ for all $n\geq n_0$.  However, $p_k(n)\to\infty$ for
  each $k\in\logic$ as $n\to\infty$.
\end{proposition}
\begin{IEEEproof}
  Since $C(\gmat)<1$ is a necessary and sufficient condition for the existence
  of a unique fixed point $\bar{\ve{p}}>0$ such that
  $\bar{\ve{p}}=\intfun(\bar{\ve{p}})$, an admission of all users to the
  network follows from Proposition \ref{prop:ACL_Totallyadmissible}. However,
  once all the users are admitted, it follows from \cite{Yates95a} and that
  fact that the SIR targets $\delta\gmat$ are not feasible (due to
  $C(\delta\gmat)\geq 1$) that the algorithm
  (\ref{eq:ActiveLinkProtection_ACL_Alg}) with $\emenge{B}_n=\emptyset$
  diverges in the sense that each transmit power tends to infinity. 
\end{IEEEproof}

\section{Incorporating power constraints}
\label{sec:RemarkPowerConstraints}

It is important to emphasize that all the properties and, in particular, the
protection of active users (ALP property) have been obtained under the
assumption of no constraints on transmit powers. Since the power at which
users transmit their signals is always limited, scepticism may arise about the
practical value of the results. This section deals with the problem under
which additional conditions the results obtained in the previous subsection
apply to power-constrained control schemes.

Everything is defined as in the previous section except that $\ve{p}\geq 0$ is
confined to be a member of some compact, convex and downward-comprehensive set
$\pset\subset\RN^K$ with $\ve{0}\in\pset$, which represents some power
constraints. The algorithm (\ref{eq:ActiveLinkProtection_ACL_Alg}) or,
equivalently, (\ref{eq:AlgwithJn}) with (\ref{eq:Jn_Def}) must be modified to
take into account these power constraints. This modification may involve the
inclusion of projection of power updates on the set $\pset$. For simplicity,
throughout this section, we assume the individual power constraints on each
link so that $\pset=\{\ve{p}\in\RN^K:\forall_{k\in\logic}p_k\leq\hat{p}_k\}$
for some given $\hat{\ve{p}}:=(\hat{p}_1,\dotsc,\hat{p}_K)>0$. In this case, a
power-constrained version of (\ref{eq:AlgwithJn}) is
\begin{equation}
\label{eq:AlgwithJn_PowerConstraints}
\ve{p}(n+1)=\delta\aclfun(\ve{p}(n),\hat{\ve{p}}/\delta),\;\ve{p}(0)\in\RP^K 
\end{equation}
where $\aclfun:\RN^{K}\times\RP^K\to\RN^{K}$ is of the form
\begin{equation}
  \label{eq:AclFunDefWithConstraints}
  \aclfun(\ve{p},\hat{\ve{p}})=\min\bigl\{\ve{p},\intfun(\ve{p}),\hat{\ve{p}}\bigr\}
\end{equation}
where the minimum is taken component-wise.

In the unconstrained case, the notions of admissibility and
$\delta$-compatibility play crucial roles for the behavior of the control
scheme (see \ref{as:fullyadmissible}--\ref{as:totallyinadmissible}). In the
presence of power constraints, however, the lack of $\delta$-compatibility has
different implications. To see this, note that (\ref{eq:FeasibilityCond}) is
necessary but not sufficient for the SIR targets to be feasible under power
constraints. A necessary and sufficient condition for feasibility of $\gmat$
is that \cite{StBookSpringer08}
\begin{equation}
\label{eq:FeasibilityCond_Constraints}
0<C(\gmat;\pset):=\min_{\ve{p}\in\pset} \max_{k\in\logic}\frac{\gamma_k
  I_k(\ve{p})}{p_k}\leq 1\,.
\end{equation}
Let $\ve{p}'\in\pset$ denote any minimizer in
(\ref{eq:FeasibilityCond_Constraints}) so that
\begin{equation}
\label{eq:FeasibilityOptPowre_Constraints}
\ve{p}':=\arg\min_{\ve{p}\in\pset}\max_{k\in\logic}\frac{\gamma_k
  I_k(\ve{p})}{p_k}\,.
\end{equation}
Obviously, as $\pset\subset\RN^K$, we have $C(\gmat)\leq C(\gmat;\pset)$, and
thus $C(\gmat)<1$ does not necessarily imply $C(\gmat;\pset)\leq 1$.  In such
cases, $C(\gmat;\pset)$ defined by (\ref{eq:FeasibilityCond_Constraints})
provides a basis for defining the notion of admissibility.  In analogy to the
previous definitions, we can say that the inactive users are
\begin{assumptions}
\item \label{as:fullyadmissible_PC}
fully admissible if $C(\gmat;\pset)\leq C(\delta\gmat;\pset)\leq 1$,
\item \label{as:deltaincomp_PC}
fully admissible but $\delta$-incompatible if $C(\gmat;\pset)\leq
  1<C(\delta\gmat;\pset)$,
\item \label{as:totallyinadmissible_PC}
totally inadmissible if $C(\gmat;\pset)>1$. 
\end{assumptions}

\begin{proposition}
\label{prop:ResultsUnderConstr}
As $n\to\infty$, $\sir_k(n)\to\bar{\sir}_k\in(0,\infty)$ and
$p_k(n)\to\bar{p}_k\in(0,\hat{p}_k],k\in\logic$, under
(\ref{eq:AlgwithJn_PowerConstraints}). If \ref{as:fullyadmissible_PC} holds,
then $\bar{\ve{p}}=\ve{p}^\circ$ where $\ve{p}^\circ>0$ is the unique vector
satisfying
\begin{equation}
\label{eq:pdeltaMinimum_def}
\ve{p}^\circ=\delta\intfun(\ve{p}^\circ)\leq\hat{\ve{p}}\,.
\end{equation}
\end{proposition}
\begin{IEEEproof}
The proof is deferred to Appendix \ref{app:proofResultsUnderConstr}.
\end{IEEEproof}

Thus, the algorithm (\ref{eq:AlgwithJn_PowerConstraints}) converges to the
fixed point of 
$\hat{\intfun}(\ve{p})=\min\bigl\{\delta\intfun(\ve{p}), \hat{\ve{p}}\bigr\}$ 
(see also the proof of the proposition), 
which is a valid power allocation provided
that \ref{as:fullyadmissible_PC} is fulfilled. This fixed point however is not
necessarily a valid power allocation if the users are fully admissible but
$\delta$-incompatible (\ref{as:deltaincomp_PC}), which stands in clear
contrast to the unconstrained case. Thus, Condition
\ref{as:fullyadmissible_PC} is crucial for the algorithm to be of any value,
which also shows that $\delta$ should be chosen very carefully. An open
question that remains is to what extent the ALP property is preserved under
\ref{as:fullyadmissible_PC} when limitations on transmit powers are taken into
account. We address this problem in the remainder of this section. From
\cite{Yates95a,Bambos00}, we know that the property of protecting active users
does not carry over in its full generality to the power-constrained case.

Considering the modified iteration (\ref{eq:AlgwithJn_PowerConstraints}) with
(\ref{eq:AclFunDefWithConstraints}) shows, together with $A2$ and $A3$, that,
for every $n\in\NNZ$ and $k\in\logic$,
\begin{equation}
\label{eq:DecreaseInterference_PC}
\begin{split}
I_k(\ve{p}(n+1))
&=I_k(\delta\aclfun(\ve{p}(n),\hat{\ve{p}}/\delta))
\switchModes{}{\\&}=I_k(\delta\min\{\ve{p}(n),\intfun(\ve{p}(n)),\hat{\ve{p}}/\delta\})\\
&<\delta I_k(\min\{\ve{p}(n),\intfun(\ve{p}(n)),\hat{\ve{p}}/\delta\})\,.
\end{split}
\end{equation}
Consequently, Lemma \ref{lem:DecreaseInterference} holds for the
power-constrained case as well, and hence, for any $n\in\NNZ$ and
$k\in\asetf_n$, one has
\begin{align}
\label{eq:SIR(n+1)withequality}
\sir_k(\ve{p}(n+1))
    &\geq\frac{\min\{\hat{p}_k,\delta\gamma_kI_k(\ve{p}(n))\}}
    {I_k(\delta\min\{\ve{p}(n),\intfun(\ve{p}(n)),\hat{\ve{p}}/\delta\})}\\
\label{eq:SIR(n+1)with1inequality}
&>\frac{\min\{\hat{p}_k,\delta\gamma_kI_k(\ve{p}(n))\}}
    {\delta I_k(\min\{\ve{p}(n),\intfun(\ve{p}(n)),\hat{\ve{p}}/\delta\})}
\end{align}
where in the first step we used the fact that
$\delta\gamma_kI_k(\ve{p}(n))\leq\delta p_k(n)$ when user $k$ is active at
time $n$. 
The following proposition shows a sufficient condition for the 
protection of active users to hold at all times.
\begin{proposition}
\label{prop:FullyAdmissiblePowerConstraints}
Let $\delta>1, n\in\NNZ$, and $k\in\asetf_n$ be arbitrary. If
$\hat{p}_k\geq\intfunk_k(\hat{\ve{p}})=\gamma_kI_k(\hat{\ve{p}})$, then we
have $k\in\asetf_{n+1}$. Thus, if $\hat{\ve{p}}$ is a valid power allocation,
that is, if
\begin{equation}
  \label{eq:hat_p_valid}
  \hat{\ve{p}}\geq\intfun(\hat{\ve{p}})
\end{equation}
then $\asetf_n\subseteq\asetf_{n+1}$ for all $n\in\NNZ$.
\end{proposition}

\begin{IEEEproof}
We consider two cases depending on whether the power constraint of an 
active user at time $n+1$ is violated or not. First assume
$\hat{p}_k\geq\delta\gamma_kI_k(\ve{p}(n))$.
An examination of (\ref{eq:SIR(n+1)with1inequality}) shows that
$\sir_k(\ve{p}(n+1))>\gamma_k$. 
Now assuming that $\hat{p}_k<\delta\gamma_k I_k(\ve{p}(n)),k\in\asetf_n$,
we see from (\ref{eq:SIR(n+1)withequality}) 
together with \eqref{eq:hat_p_valid}
that $\sir_k(\ve{p}(n+1))$ is
bounded below by $\hat{p}_k/I_k(\hat{\ve{p}})$. This completes the proof.
\end{IEEEproof}

Note that if \ref{as:fullyadmissible_PC} holds, then (\ref{eq:hat_p_valid}) is
implied by $\delta\intfun(\ve{p}')\geq\intfun(\hat{\ve{p}})$, which in turn is
satisfied if $\delta\ve{p}'\geq\hat{\ve{p}}$, where $\ve{p}'$ is defined by
(\ref{eq:FeasibilityOptPowre_Constraints}). This is simply because if
\ref{as:fullyadmissible_PC} is true, we have
$\delta\intfun(\ve{p}')\leq\ve{p}'\leq\hat{\ve{p}}$. As
$\ve{p}^\circ\leq\ve{p}'\leq\hat{\ve{p}}$, it is obvious that
$\asetf_n\subseteq\asetf_{n+1}$ for all $n\in\NNZ$ whenever
$\delta\intfun(\ve{p}^\circ)\geq\intfun(\hat{\ve{p}})$ or
$\delta\ve{p}^\circ\geq\hat{\ve{p}}$, where $\ve{p}^\circ$ is given by
(\ref{eq:pdeltaMinimum_def}) and is independent of $\hat{\ve{p}}$. All these
conditions are more restrictive than (\ref{eq:hat_p_valid}) but they may be of
interest when for instance $I_k(\hat{\ve{p}})$ is not known or difficult to
determine.

\subsection{Distress Signaling}
\label{sec:RemarkPowerConstraints_distress}

The foregoing conditions are independent of $n\in\NNZ$, and thus, if they are
satisfied, the ALP property is guaranteed for all $n\in\NNZ$, just as in the
case of unconstrained transmit powers. Now the question is what to do when
(\ref{eq:hat_p_valid}) and with it all the consequential conditions, cannot be
guaranteed. One possible remedy is to apply the concept of distress signaling
where users are prohibited from increasing their transmit powers
whenever they receive a distress signal (special tone in a control slot or
some separate control channel) broadcasted by at least one active user. The
idea was already mentioned in \cite{Bambos00} where the distress signal is
suggested to be broadcasted when an active user is about to exceed its power
limit at some time point, that is, when
$\hat{p}_k<\delta\intfunk_k(\ve{p}(n))$ for some $k\in\logic$ and
$n\in\NNZ$. One problem with this approach is that the active users may be
about to violate their power constraints again and again, thereby generating
distress signals at many different time points. In some situations, it would
be better not to deactivate the distress signal until it is guaranteed that
all the inactive users can be admitted with the protection of active users.

In this subsection, we derive more general conditions under the assumption of
standard interference functions. First we slightly strengthen the condition
$\delta\ve{p}'\geq\hat{\ve{p}}$.

\begin{proposition}
\label{prop:pn<delta*p^ast}
Suppose that \ref{as:fullyadmissible_PC} is satisfied and $\ve{p}$ is any
power vector such that 
\begin{equation}
\label{eq:delta_valid_powervector}
\delta\intfun(\ve{p})\leq\ve{p}\leq\hat{\ve{p}}\,.
\end{equation}
Let $\lambda:=\lambda(\delta,\ve{p})$ be any constant for which
$\intfun(\lambda\delta\ve{p})\leq\hat{\ve{p}}$. If
\begin{align}
\label{eq:pn<delta*p^ast}
\ve{p}(m)\leq\lambda\delta\ve{p}
\end{align}
for some $m\in\NNZ$, then $\asetf_{n}\subseteq\asetf_{n+1}$ for all $n\geq m$.
\end{proposition}
\begin{IEEEproof}
  We refer to Appendix \ref{app:proofpn<delta*p^ast}. 
\end{IEEEproof}

Any power vector satisfying (\ref{eq:delta_valid_powervector}) is called a
\emph{$\delta$-valid power vector/allocation}. Notice that if
\ref{as:fullyadmissible_PC} hold, a $\delta$-valid power vector exists.
Particular examples of such vectors are $\ve{p}'$ and $\ve{p}^\circ$ defined
by (\ref{eq:FeasibilityOptPowre_Constraints}) and
(\ref{eq:pdeltaMinimum_def}), respectively. Also note that due to $A2$ and
Proposition \ref{prop:InterFunContinuous}, there exists $\lambda$ strictly 
larger than $1$.

By Proposition \ref{prop:pn<delta*p^ast}, we have the ALP property if the
inactive users are totally admissible and the transmit powers are sufficiently
small so that (\ref{eq:pn<delta*p^ast}) is fulfilled. A useful property of
this result is that once (\ref{eq:pn<delta*p^ast}) is satisfied, there is no
need to verify this condition again, unless (\ref{eq:delta_valid_powervector})
is violated due to, for instance, fading effects or arrival of new inactive
users. The main problem with (\ref{eq:pn<delta*p^ast}), however, is how to
efficiently obtain a $\delta$-valid power allocation in a distributed
environment. One possibility is to bound below the set of $\delta$-valid
power allocations under the worst-case scenario. This problem is left
open. Instead we consider the possibility of letting each user compare its
transmit power with the interference power. First, we use Proposition
\ref{prop:pn<delta*p^ast} to prove the following result.

\begin{proposition}
\label{prop:pn<deltaIn_protection}
Assume \ref{as:fullyadmissible_PC} and let $\lambda\geq 1$ be defined as in
Proposition \ref{prop:pn<delta*p^ast}. If
\begin{equation}
\label{eq:pn<lambda_deltaIn/lambda}
\frac{\ve{p}(m)}{\lambda\delta}
\leq\delta\intfun\Bigl(\frac{\ve{p}(m)}{\lambda\delta}\Bigr)
\end{equation}
for some $m\in\NNZ$, then $\asetf_n\subseteq\asetf_{n+1}$ for all $n\geq m$.
\end{proposition}
\begin{IEEEproof}
  See Appendix \ref{app:pn<deltaIn_protection_proof}. 
\end{IEEEproof}

Notice that by Proposition \ref{prop:InterFunContinuous}, $A2$ and
(\ref{eq:pdeltaMinimum_def}), there exists $\lambda>1$ satisfying the
condition of the proposition: $\intfun(\lambda\delta\ve{p}^\circ)
\leq\hat{\ve{p}}$. Choosing $\lambda=1$ leads us to the following corollary.
\begin{corollary}
\label{cor:pn<delta2In_protection}
If \ref{as:fullyadmissible_PC} holds and
\begin{equation}
\label{eq:pn<delta2In/lambda}
\ve{p}(m)\leq\delta^2\intfun\bigl(\ve{p}(m)/\delta)
\end{equation}
for some $m\in\NNZ$, then $\asetf_n\subseteq\asetf_{n+1}$ for all $n\geq m$. 
\end{corollary}

We point out that it is not clear whether (\ref{eq:pn<lambda_deltaIn/lambda}),
and with it (\ref{eq:pn<delta2In/lambda}), is preserved in general. The
results solely show that once (\ref{eq:pn<lambda_deltaIn/lambda}) or
(\ref{eq:pn<delta2In/lambda}) is satisfied, then the ALP property is ensured
for all time instances $n\geq m$. It must be also emphasized that
(\ref{eq:pn<lambda_deltaIn/lambda}) and (\ref{eq:pn<delta2In/lambda}) are
\emph{less restrictive} than $\ve{p}(m)\leq\delta\intfun(\ve{p}(m))$ as
$x\mapsto x\intfunk_k(\ve{p}/x)$ is strictly increasing for any $\ve{p}>0$
(see also the proof of Proposition \ref{prop:pn<delta2In_protection_always}).

The main problem with Proposition \ref{prop:pn<deltaIn_protection} and
Corollary \ref{cor:pn<delta2In_protection} is that
$\intfunk_k(\ve{p}(m)/(\lambda\delta))$ and $\intfunk_k(\ve{p}(m)/\delta)$ may
be not known to user $k$ at time $m$ even if $\intfunk_k(\ve{p}(m))$ is
known.\footnote{In some cases, e.g. when $I_k$ is a linear interference function
  (\ref{eq:LinearInterFun}) and the noise factor is known, the
  value $\intfunk_k(\ve{p}(m)/\delta)$ can be obtained from
  $\intfunk_k(\ve{p}(m))$} The following proposition shows that the ALP
property is guaranteed even if $\ve{p}(m)>\delta\intfun(\ve{p}(m))$, provided
that the entries of $\ve{p}(m)$ are not too large. In other words, there is
always some margin around the value $\delta\intfun(\ve{p}(m))$ so that the
protection is guaranteed whenever $\ve{p}(m)$ belongs to this margin.

\begin{proposition}
\label{prop:pn<delta2In_protection_always}
Suppose that \ref{as:fullyadmissible_PC} is true and
\begin{equation}
\label{eq:pn<delta2In/lambda_always}
\ve{p}(m)\leq\beta\delta\intfun\bigl(\ve{p}(m))
\end{equation}
holds for some $m\in\NNZ$ and $\beta\in[1,\beta_{\max}]$. Then, there exists
$\beta_{\max}>1$ such that $\asetf_n\subseteq\asetf_{n+1}$ for all $n\geq m$.
\end{proposition}
\begin{IEEEproof}
  See Appendix \ref{app:pn<delta2In_protection_always_proof}. 
\end{IEEEproof}

By the proposition, we have the protection of active users for all $n\geq m$ if
(\ref{eq:pn<delta2In/lambda_always}) holds for some sufficiently small
$\beta\geq 1$. The main insight is that there is the possibility of choosing
$\beta$ being \emph{strictly} larger than one.


	In the remainder of this section, we summarize our findings and make
  some suggestions as to what to do when \ref{as:fullyadmissible_PC} is not
  fulfilled. For brevity, we focus on condition
  \eqref{eq:pn<delta2In/lambda_always} but the subsequent discussion also
  applies to \eqref{eq:pn<lambda_deltaIn/lambda} and
  \eqref{eq:pn<delta2In/lambda} (with \eqref{eq:pn<delta2In/lambda_always}
  substituted by \eqref{eq:pn<lambda_deltaIn/lambda} or
  \eqref{eq:pn<delta2In/lambda}). Given some $\beta>1$ and $\delta>1$ (both
  sufficiently small), let $\pset'\subseteq\pset$ be the set of all power
  allocations for which \eqref{eq:pn<delta2In/lambda_always} is
  satisfied. When $\ve{p}(n)\notin\pset'$, the scheme prevents all users from
  increasing their powers by broadcasting distress signals on a common control
  channel. The distress signals are sent by all the active users $k\in\aset_n$
  such that $p_k(n)>\beta\delta\intfunk_k(\ve{p}(n))$, which can be verified
  locally.  First assume that \ref{as:fullyadmissible_PC} holds, meaning that
  there is an additional mechanism to ensure full admissibility of all
  users. Then, the admission control algorithm with distress signaling
  becomes:
  \begin{equation}
    \label{eq:ACL_DistressSignaling}
    \ve{p}(n+1)=\begin{cases}
      \min\bigl\{\ve{p}(n), \delta\intfun(\ve{p}(n))\bigr\} &
      \ve{p}(n)\notin\pset'\\
      \delta\aclfun(\ve{p}(n),\hat{\ve{p}}/\delta) &\ve{p}(n)\in\pset'
      \end{cases}
  \end{equation}
  where $\aclfun$ is defined by (\ref{eq:AclFunDefWithConstraints}).  From
  (\ref{eq:ACL_DistressSignaling}), we see that the admission control
  algorithm (\ref{eq:AlgwithJn_PowerConstraints}) stops if
  $\ve{p}(n)\notin\pset'$ (at least one active user transmits a distress
  signal), in which case no user increases its transmit power. Therefore,
  active users are protected as the interference powers do not increase and
  each active user, say user $k\in\aset_n$, decreases its transmits power if
  and only if $p_k(n)>\delta\intfunk_k(\ve{p}(n))$. Moreover, since the
  transmit power of user $k$ decreases as long as
  $p_k(n)>\delta\intfunk_k(\ve{p}(n))$ and other transmit powers are kept
  constant, there must be a time point $m\geq n$ such that
  $\ve{p}(m)\in\pset'$. Once this condition is satisfied, no distress signal
  is broadcasted and, by (\ref{eq:ACL_DistressSignaling}), the iteration
  \eqref{eq:AlgwithJn_PowerConstraints} is resumed. Now the active users are
  guaranteed to be protected for all $n\geq m$, provided that
  \ref{as:fullyadmissible_PC} is satisfied. 

  Now if \ref{as:fullyadmissible_PC} is not satisfied, the problem is open but
  we have to differentiate between \ref{as:deltaincomp_PC} and
  \ref{as:totallyinadmissible_PC}. In the case of \ref{as:deltaincomp_PC}, the
  algorithm in (\ref{eq:ACL_DistressSignaling}) applies, provided that the
  parameter $\delta>1$ is reduced so as to fulfill \ref{as:fullyadmissible_PC}
  at the expense of extending the duration of the whole admission process. So
  the only issue is when and how to reduce $\delta$ to provide full
  admissibility. In contrast, if \ref{as:totallyinadmissible_PC} is true, then
  it is impossible to admit all users at the required quality-of-service, and
  therefore the SIR target of some active user will be violated at some time
  point. A simple idea is then to let this active user permanently send a
  distress signal so that no transmit powers are increased and, after some
  time point, first inactive users will drop out of the system. Obviously, a
  better approach would be to let inactive user (cooperatively) estimate
  $C(\gmat;\pset)$ and $C(\delta\gmat;\pset)$ so that they do not even attempt
  to access the network if $C(\gmat;\pset)>1$. However, an efficient
  estimation of these quantities in a distributed environment is still an open
  problem.
  
  We point out that there are two possible interpretations of the admission problem
  and the scheme \eqref{eq:ACL_DistressSignaling}. The first one is \emph{how} to 
  admit inactive users if we know that they are fully admissible in the sense 
  of \ref{as:fullyadmissible_PC}, and the second one is \emph{whether} to
  admit them to the network when we do not know if they are admissible or not.
  In the former case, \eqref{eq:ACL_DistressSignaling}
  guarantees that the incoming users will be admitted in finite time and the 
  existing ones will be protected. In the latter case, the incoming users
  will either be admitted (as described above), or the SIR target of some active user
  will be violated for a single time point, which will prove that the condition
  \ref{as:fullyadmissible_PC} is not satisfied. Therefore, the considered
  scheme does not lead to admission errors and the decision is always made within 
  finite time. However, the exact number of iterations of \eqref{eq:ACL_DistressSignaling}
  needed to make a decision is not determined. Especially if the SIR targets
  are close to the maximum supportable SIR targets in the system this number may be high
  which may lead to noticeable delays.

\section{Linear transceiver optimization}
\label{sec:ActiveLinkProtection_Beamforming}

Any user intending to access the network must select its transmit vector by
determining, for instance, its beamforming vector. The transmit vectors have
different physical meanings, depending on the realization of the physical
layer. Abstractly speaking, the transmit vectors determine the ``directions''
of transmit signals in some appropriately chosen signal space (see also the
multiple antenna case below). Note that once the transmit vectors of the
inactive users are determined, they cannot be modified arbitrarily as the
iteration \eqref{eq:ActiveLinkProtection_ACL_Alg} does not guarantee the
protection of active users under such modifications. On the other hand, the
transmit vectors of active users may prevent an inactive user from entering
the network.
Inspired by \cite{UlukusWireless04}, we alleviate this problem by considering
a scheme in which all transmit vectors (and the power vector) are recalculated
so as not to deteriorate the SIR performance of the users. During the
transmitter side optimization, the execution of the admission control
iteration \eqref{eq:ActiveLinkProtection_ACL_Alg} is suspended. Our
  objective in this section is to show how much performance gains can be
  expected by optimizing transmit vectors in addition to power control and
  receiver-side optimization. Note that due to an optimization of transmit
  vectors, the scheme does not fall within the framework presented in the
  previous sections.

The basic idea is to carry out an iterative optimization of transmit and
receive vectors\footnote{For instance, the reader can think of transmit and
  receive vectors as transmit and receive beamformers, respectively. Notice
  that there is no more than one data stream per link.} in an alternating
manner, with receivers and transmitters exchanging their roles. The transmit
beamformers are optimized in the \emph{reversed network}, which is the network
obtained by reversing the direction of all links and exchanging the roles of
transmitters and receivers on each link (with the actual transmit vectors used
as receive vectors).  We leverage the fact \cite{UlukusWireless04} that if any
given SIR values are feasible in the primal network, they are also feasible in
the reversed network (albeit with a different power
allocation).
Every iteration of the algorithm consists of two steps in the primal network
and two steps in the reversed network. These two steps to be performed in each
network are receive beamformer optimization and power vector
computation. Given fixed transmit beamformers and power allocation, the
receiver-side optimization in the sense of (\ref{eq:InterFunMMSE}) can be
performed in a distributed manner using either pilot-based or blind estimation
methods \cite{Ve98}. The power allocations in the primal network can
  be computed in a decentralized manner using the distributed asynchronous
  on-line power control algorithm of \cite{Foschini93,Yates95a} with the SIR
  values from the reversed network treated as the SIR targets.  The power
  allocation for the reversed network can be computed in the same fashion with
  the SIR values from the primal network.
  For more details about the considered transceiver optimization scheme 
  the reader is referred to \cite{StKaliszAsil08}.

The transceiver optimization scheme discussed in this section is supposed to
be performed on a regular basis (periodically) or only when necessary. 
The distributed implementation of the entire scheme, consisting of admission/power
control and transceiver optimization, is more challenging compared
to the pure admission control scheme presented in Sections 
\ref{sec:ActiveLinkProtection_ACL}-\ref{sec:RemarkPowerConstraints}.
Previously, only local interference measurements and a common signaling channel to
broadcast distress signals were required; now, the users are supposed to be
willing to suspend their normal operation and coordinately agree to 
enter a different operation mode in which they jointly 
optimize transmit powers and beamformers. However, due to potentially very
high performance gains (see simulations in the next section) it seems
encouraging to consider at least some reduced form of transceiver
optimization in future research.

\begin{figure}[t]
  \begin{center}
\subfloat[SIR behavior]{\label{fig:sirs}
\scalebox{0.6}{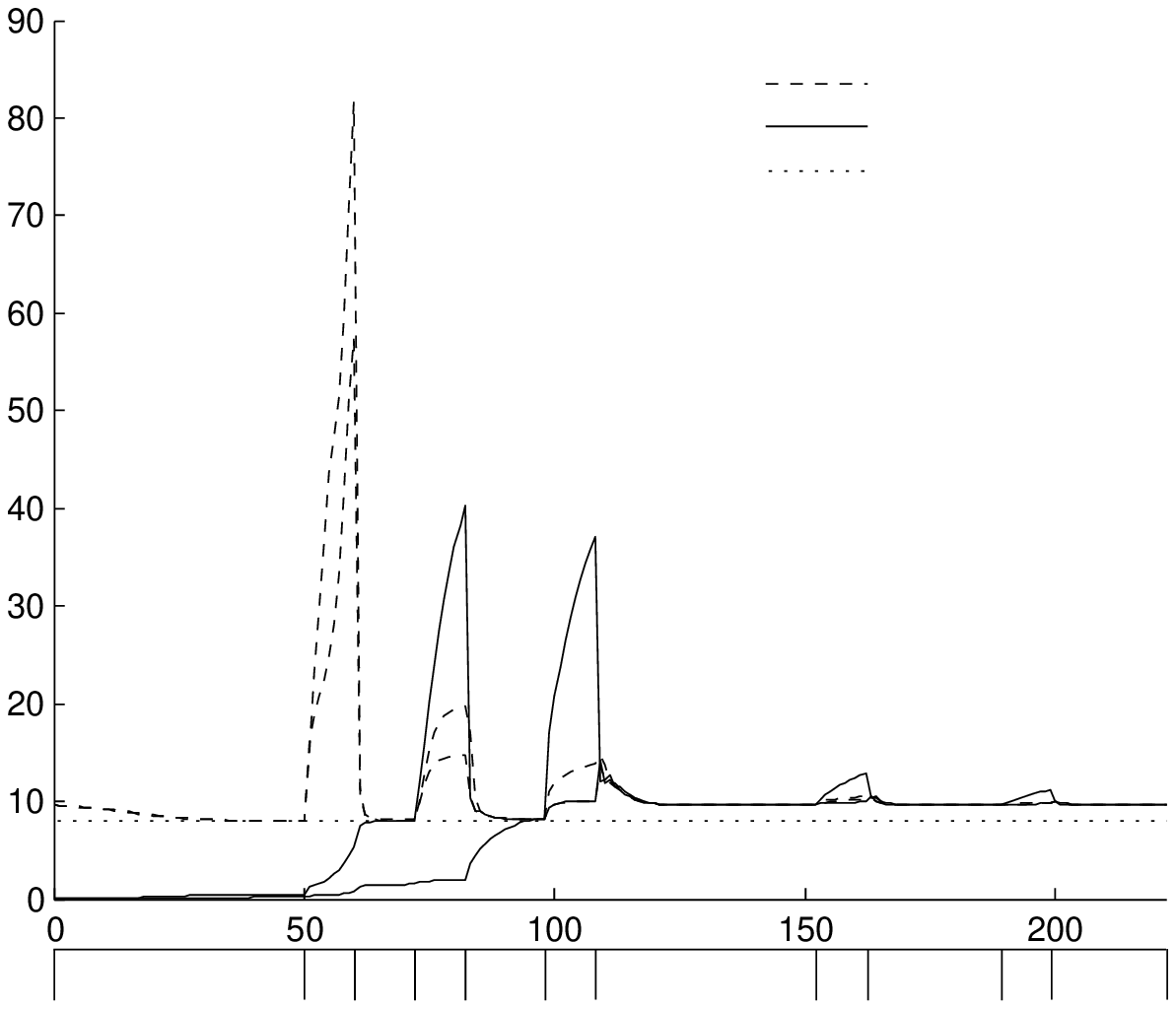}
}
\switchModes{}{\\}
\subfloat[Transmit powers]{	\label{fig:tx_powers}
\scalebox{0.6}{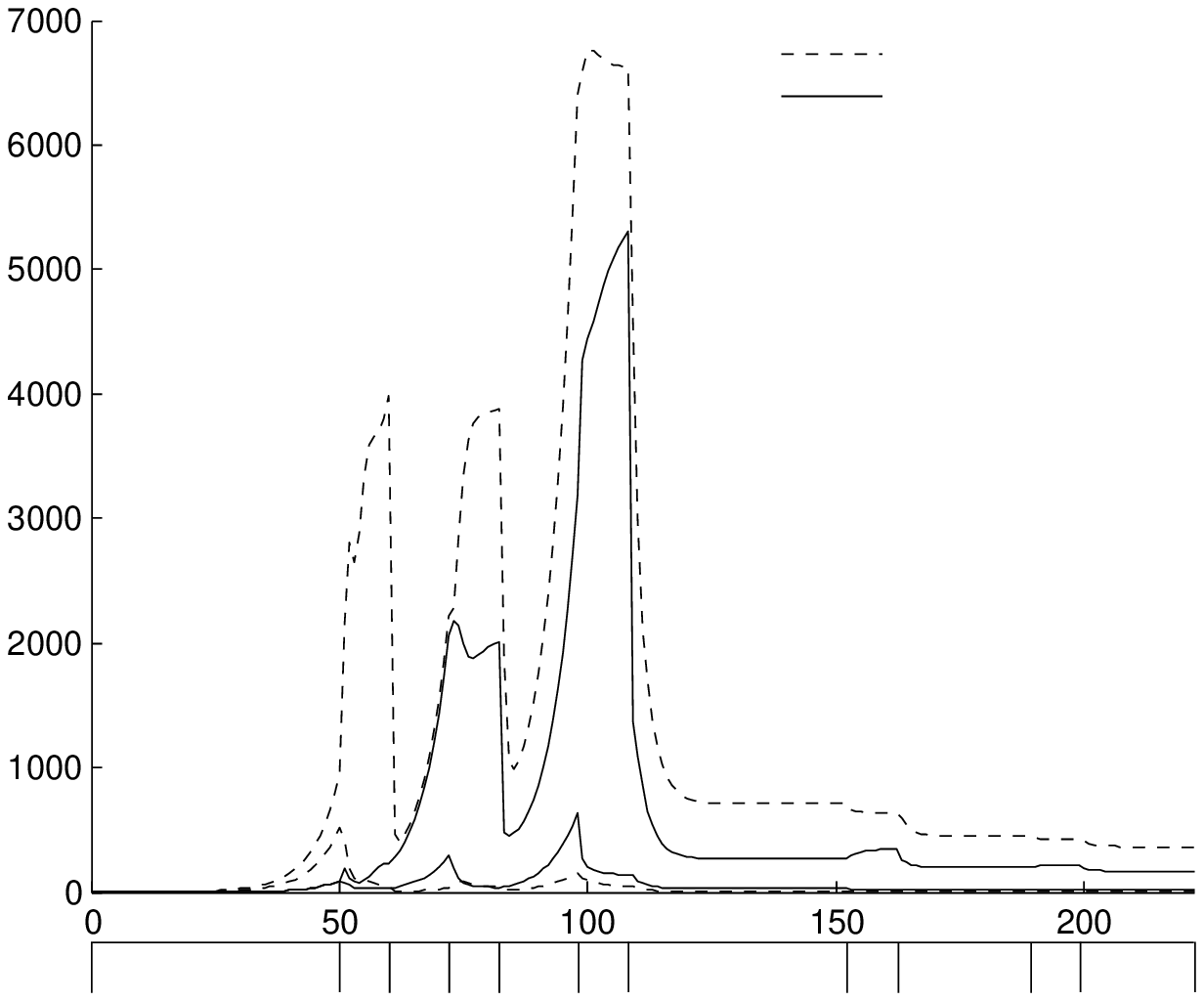}
}
\caption{The horizontal axis in both figures represents the iterations,
        and the legend below the axis provides the information whether these
        are the iterations of: A - admission and power control
        \eqref{eq:ActiveLinkProtection_ACL_Alg}, or T - transceiver
        optimization. 
        }
\label{fig:optimization_all}
\end{center}
\end{figure}
\begin{figure}[t]
  \begin{center}
  \subfloat[Admission Control - phase A.1 in Fig. \ref{fig:optimization_all}]{
  	\label{fig:ac_1}
		\scalebox{0.5}{\inputsinrsubfloat{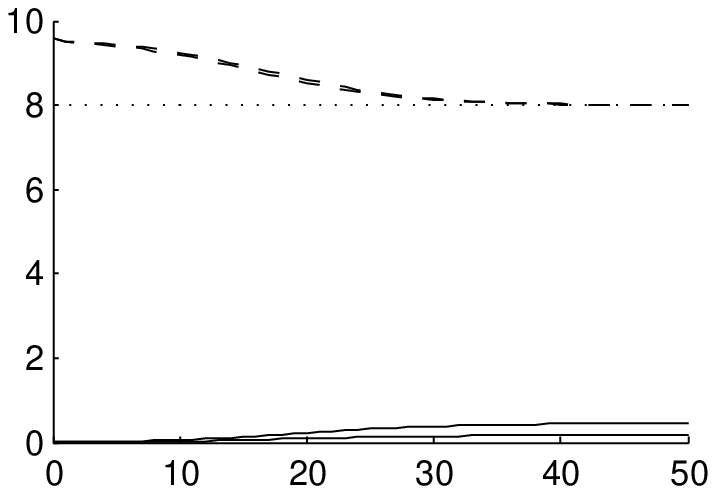}}
	\scalebox{0.5}{\inputtxpowersubfloat{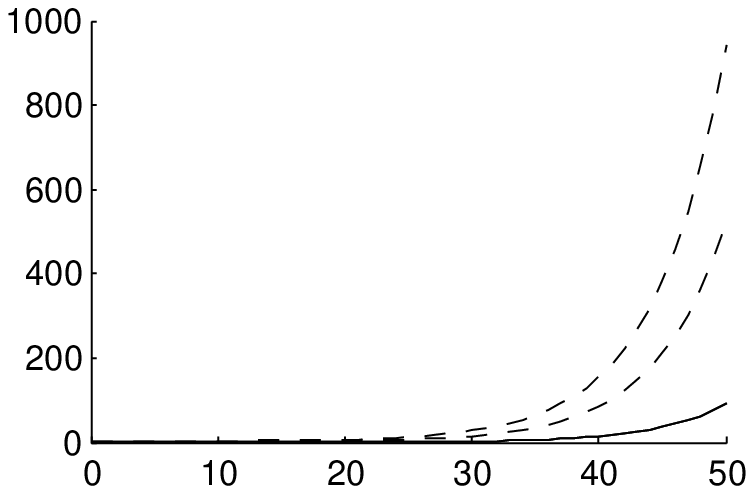}}}
	\switchModes{}{\\}
	\subfloat[Admission Control - phase A.2 in Fig. \ref{fig:optimization_all}]{
		\label{fig:ac_2}
	\scalebox{0.5}{\inputsinrsubfloat{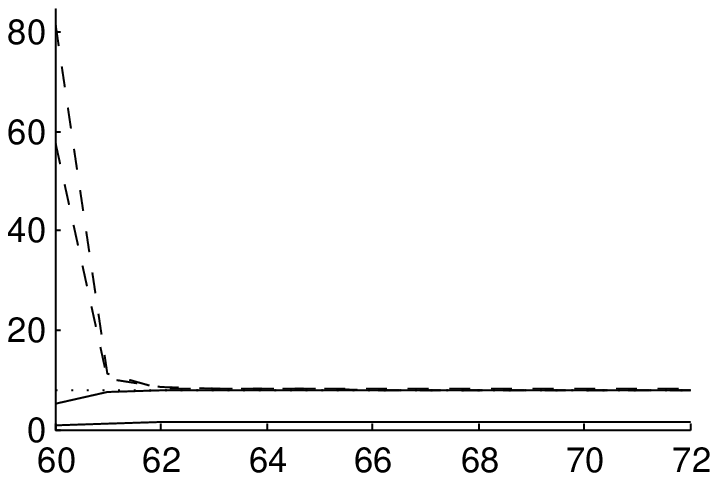}}
	\scalebox{0.5}{\inputtxpowersubfloat{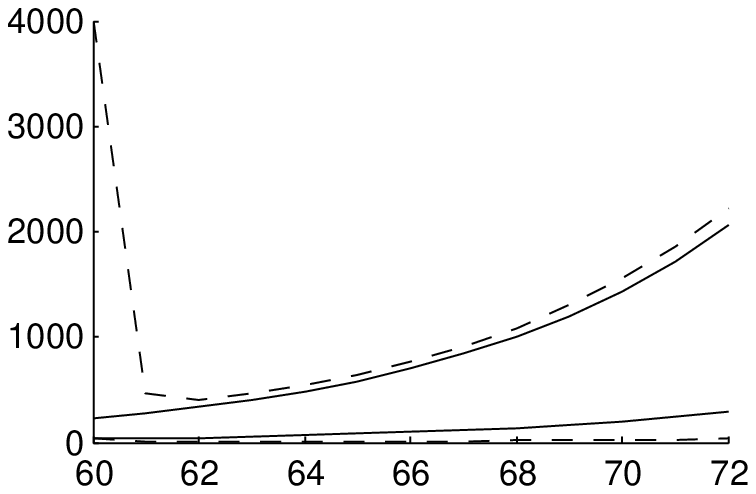}}}\\
	\subfloat[Admission Control - phase A.3 in Fig. \ref{fig:optimization_all}]{
		\label{fig:ac_3}
	\scalebox{0.5}{\inputsinrsubfloat{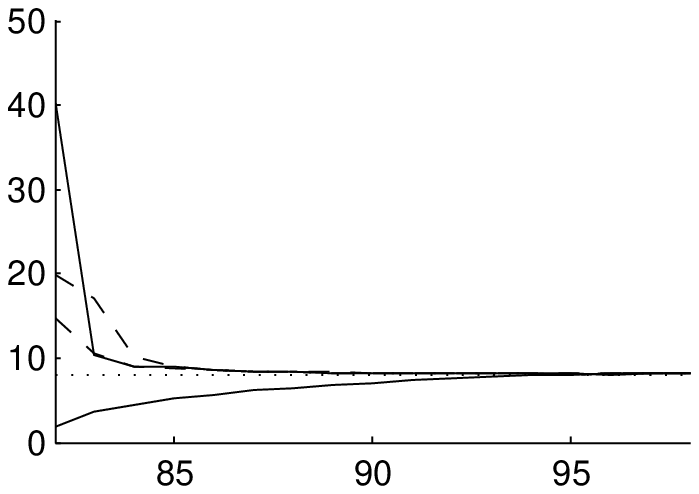}}
	\scalebox{0.5}{\inputtxpowersubfloat{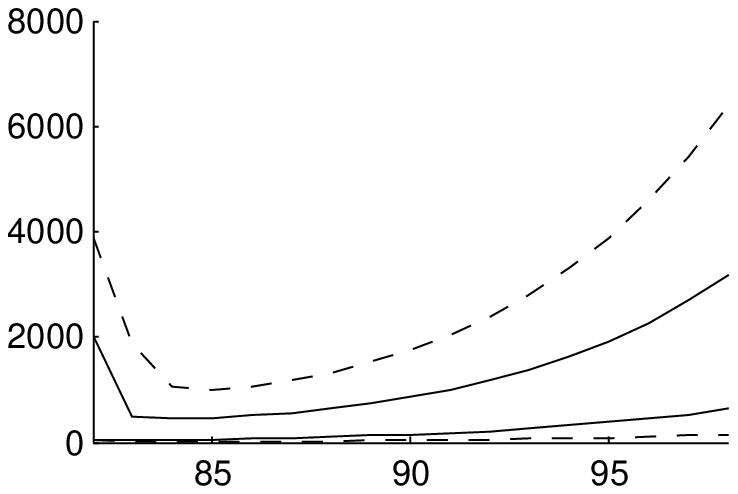}}}
	\switchModes{}{\\}
	\subfloat[Admission Control - phase A.4 in Fig. \ref{fig:optimization_all}]{
		\label{fig:ac_4}
	\scalebox{0.5}{\inputsinrsubfloat{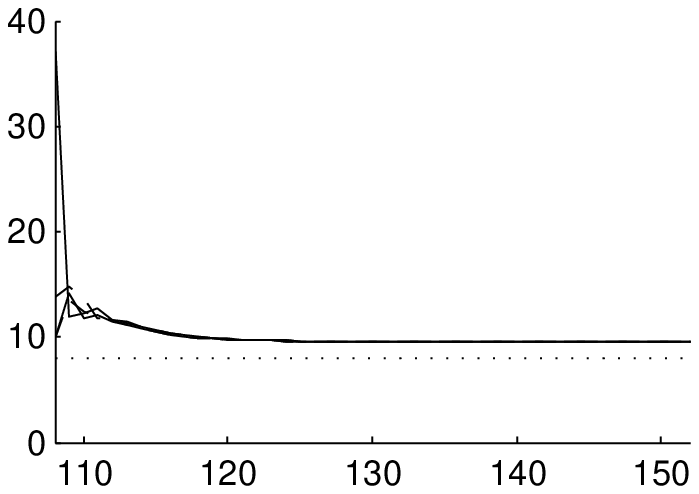}}
	\scalebox{0.5}{\inputtxpowersubfloat{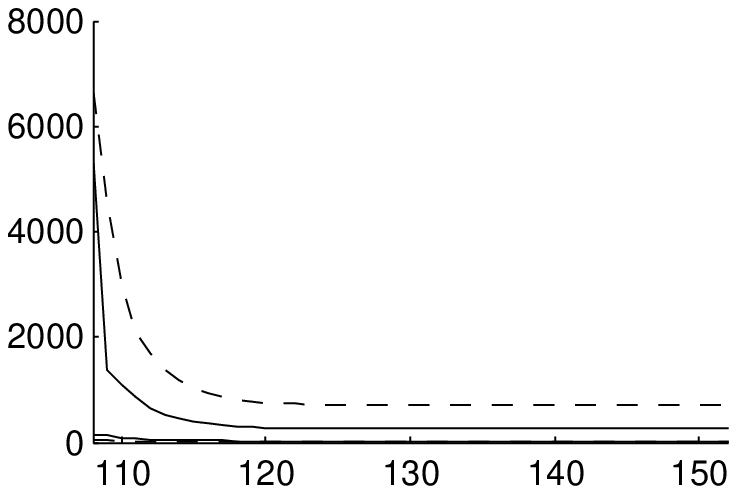}}}\\
	\subfloat[Transceiver Optimization - phase T.2 in Fig. \ref{fig:optimization_all}]{
		\label{fig:tx_1}
	\scalebox{0.5}{\inputsinrsubfloat{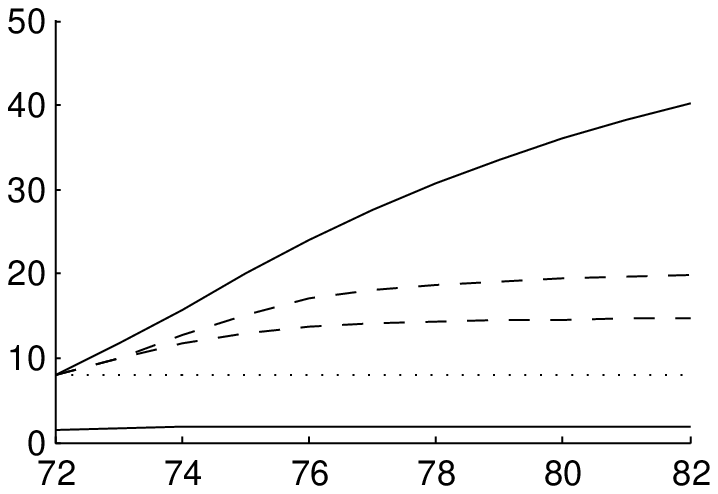}}
	\scalebox{0.5}{\inputtxpowersubfloat{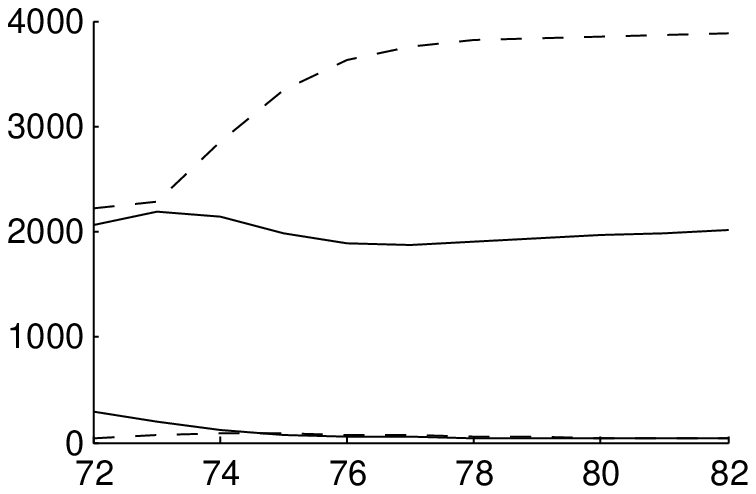}}}
\caption{Selected optimization phases}
\label{fig:optimization_phases}	

\end{center}
\end{figure}

\subsection{Simulations}
\label{sec:ActiveLinkProtection_simulation}

 We assume a wireless network with $K=10$ users and without power
  constraints, in which the receivers and transmitters are equipped with
  $n_R=4$ and $n_T=4$ antennas, respectively. The wireless channel is assumed
to be flat and fixed, with the complex-valued channel coefficients on each
link being realizations of i.i.d. circular symmetric normal distributed random
variables with zero mean and unit variance. 
For brevity, there is a common SIR target for all users $\gamma=8$. Five users
are already admitted at time $0$ with their SIRs equal to $\delta\gamma=9.6$,
and their transmit and receive beamformers are already pre-optimized (a number
of iterations of the transceiver optimization algorithm were performed for
those five users). The remaining five users are inactive at time $0$ and
transmit at some low power. The transmit and receive beamformers of the
inactive users were initialized according to the singular value decomposition
(SVD) of the corresponding channel
matrices.

Figures \ref{fig:sirs} and \ref{fig:tx_powers} present simulation results of a
wireless network, in which both admission control
\eqref{eq:ActiveLinkProtection_ACL_Alg} and transceiver optimization
 are carried out.  The
evolution of the SIRs of the individual users is presented in
Fig. \ref{fig:sirs}, whereas Fig. \ref{fig:tx_powers} depicts the
corresponding transmit powers.  For a better readability of the plots, four
representative users were selected: two initially inactive ones and two
initially active ones (shown with solid and with dashed lines,
respectively). The curves for the remaining six users are not shown, they
evolve however in a similar way to the presented ones. The simulation was
configured to start with the admission control scheme
\eqref{eq:ActiveLinkProtection_ACL_Alg}, which was executed until the SIRs of
the users converged (with some given accuracy) or until transmit powers
exceeded a given level. Subsequently, the transceiver optimization
was performed (for a fixed number of 10 iterations). Then the execution of
admission control was resumed; this cycle was repeated 5 times, and in the
6-th, final cycle, only admission control was performed.

Figure \ref{fig:optimization_phases} presents selected phases of the
optimization in more detail. In particular, Fig. \ref{fig:ac_1} shows the
first admission control phase.  As mentioned above, the transceivers of the
initially active users were pre-optimized, but those of the inactive users
were not. This leads to high interference for the inactive users, and in
consequence none of them achieves the SIR target and can be admitted to the
network in this phase, although the transmit powers tend to grow to infinity.
It can be however seen that the SIRs of the initially active users do not drop
below the target value. Afterwards, the transceiver optimization scheme is
performed (phase T.1). The subsequent admission control phase (A.2) is
depicted in Fig \ref{fig:ac_2}.  We see that one of the previously inactive
users can be admitted, but the other one still remains below the SIR
target. Thus, this phase (as well as phase A.1) corresponds to the case
\ref{as:totallyinadmissible} where the users are fully inadmissible.  Then,
there follows again a transceiver optimization phase. This phase (T.2) is presented
in more detail in Fig. \ref{fig:tx_1}. The SIRs of all users are increasing,
and it can be observed that the transmit powers do not have the tendency to
fast (geometrical) growth. In the following admission control phase A.3
(Fig. \ref{fig:ac_3}), the remaining inactive user is eventually admitted to
the network. The transmit powers still tend to grow to infinity, though, and
the SIRs converge to $\gamma$ instead of $\delta\gamma$. This indicates that
the users are fully admissible, but $\delta$-incompatible (case
\ref{as:deltaincomp}). However, after another transceiver optimization T.3 is
performed, users finally become both fully admissible and $\delta$-compatible
(case \ref{as:fullyadmissible}). This is presented in Fig. \ref{fig:ac_4} -
the SIRs converge to $\delta\gamma$ and transmit powers to some constant
values (as opposed to growing to infinity in the previous phases). The
transceiver optimization performed in the following two cycles (T.4 and T.5)
leads to significantly lowering the maximum transmit power. It should be also
noted that transmit powers in the final phase converge to much lower values
than during the intermediate phases.

In the presented simulation, all 10 users were able to reach the common SIR
target $\gamma=8$. Extensive simulations have shown that at least the SIR
target of $24$ can be achieved with finite powers, provided that a
sufficiently high number of transceiver optimization phases is performed. For
comparison, with no optimization of the transmitters, but with optimal receive
beamformers, the highest observed common SIR target attained by all users was
approximately $1.37$, and when both transmit and receive beamformers were
fixed and equal to the SVD vectors, the highest feasible SIR target is
0.88. These observations indicate a potential for significant performance
gains of the transmitter side optimization.

\section{Appendix}
\label{app:Appendix}

\subsection{Proof of Lemma \ref{lem:convGeneralFun}}
\label{app:proofLemmaAsym}

  Let $\ve{p}\geq 0$ and $k\in\logic$ be arbitrary. By $A1$,
  $\intfunk_k(c\ve{p})/c$ is positive, and hence bounded below by $0$ for all
  $c>0$. Moreover, $A2$ implies that
  $\intfunk_k(c_1\ve{p})/c_1=(c_2/c_2)\,\intfunk_k(c_1\ve{p})/c_1
  >\intfunk_k(c_2c_1/c_1\ve{p})/c_2=\intfunk_k(c_2\ve{p})/c_2$ for any
  $0<c_1<c_2$. Thus,
  $\gintfunk_k(\ve{p})=\lim_{c\to\infty}\intfunk_k(c\ve{p})/c$ exists and is
  nonnegative. This proves the existence of $\gintfunk_k$ and
  $\tilde{A}1$. The homogeneity property $\tilde{A}2$ holds since, for any
  $\mu>0$, one has
  $\gintfunk_k(\mu\ve{p}) =\lim_{c\to\infty}\intfunk_k(c\mu\ve{p})/c
  =\lim_{c\to\infty}\mu\intfunk_k(c\mu\ve{p})/(\mu c)
  =\mu\lim_{c'\to\infty}\intfunk_k(c'\ve{p})/c'
  =\mu\gintfunk_k(\ve{p})$. Finally, by $A3$, for any $\ve{p}^{(1)}\geq
  \ve{p}^{(2)}\geq 0$ and $c>0$, we have $0\leq
  \intfunk_k(c\ve{p}^{(1)})/c-\intfunk_k(c\ve{p}^{(2)})/c$. So, $0\leq
  \lim_{c\to\infty}(\intfunk_k(c\ve{p}^{(1)})/c-\intfunk_k(c\ve{p}^{(2)})/c)=\gintfunk_k(\ve{p}^{(1)})
  -\gintfunk_k(\ve{p}^{(2)})$, from which $\tilde{A}3$ follows.

\subsection{Proof of Proposition \ref{prop:ACL_TotallyInadmissible}}
\label{app:proofTotallyInAdmissible}

By \ref{as:totallyinadmissible}, the network is totally inadmissible so that
$\emenge{B}\neq\emptyset$. By Proposition \ref{prop:ACL_SIRinactiveGrows}, we
have $\sir_k(n)<\sir_k(n+1)$ for each $k\in\emenge{B}_n$. As a consequence,
there must be a (sufficiently large) number $N$ such that
$\emenge{B}=\emenge{B}_n\neq\emptyset$ and
$\emenge{A}=\emenge{A}_n\neq\emptyset$ for all $n\geq N$. Note that
$\emenge{A}\cap\emenge{B}=\emptyset$. Unless otherwise stated, assume that
$n\geq N$. Since $p_k(n+1)=\delta^{n+1}p_k(0)$ for each $k\in\emenge{B}$, we
trivially obtain $p_k(n)/\delta^n=p_k(0)=\bar{p}_k,k\in\emenge{B}$, for all
$n\in\NNZ$. Moreover, as the sequence $\{\sir_k(n)\}_{n\geq N}$ is strictly
increasing (Proposition \ref{prop:ACL_SIRinactiveGrows}) and bounded above by
$\gamma_k,k\in\emenge{B}$, it must converge to some
$\bar{\sir}_k\leq\gamma_k$. Now let us consider the transmit powers of the
active users. Defining $\ve{\lambda}=\ve{p}^{(i)}(0)>0$, we have
  \begin{align*}
    \ve{p}^{(a)}(n+1)&=\delta\intfun^{(a)}(\ve{p}(n))
    =\delta\intfun^{(a)}\begin{pmatrix}\begin{pmatrix}\ve{p}^{(a)}(n),
        \ve{p}^{(i)}(n)\end{pmatrix}\end{pmatrix}
    \switchModes{}{\\&}=\delta\intfun^{(a)}\begin{pmatrix}\begin{pmatrix}\ve{p}^{(a)}(n),
        \delta^n\ve{\lambda}\end{pmatrix}\end{pmatrix}\\
    &=\delta\intfun^{(a)}\begin{pmatrix}\delta^n\begin{pmatrix}\frac{1}{\delta^n}\ve{p}^{(a)}(n),
        \ve{\lambda}\end{pmatrix}\end{pmatrix}\,.
  \end{align*}
  Thus, using $\ve{\pi}(n)=\ve{p}^{(a)}(n)/\delta^n,n\geq N$, we can write
  \begin{align}
    \label{eq:IterationPi(n+1)=Ia(pi(n))}
    \ve{\pi}(n+1)=
    \intfun^{(a)}\begin{pmatrix}\delta^n\begin{pmatrix}\ve{\pi}(n),
        \ve{\lambda}\end{pmatrix}\end{pmatrix}/\delta^n,\;n\geq N\,.
    \end{align}
    Now suppose that $\{\ve{\pi}(n)\}_{n\geq N}$ is a sequence generated by
    (\ref{eq:IterationPi(n+1)=Ia(pi(n))}).  Thus, since
    $\sir_k(n)=\tfrac{\gamma_kp_k(n)/\delta^n}
    {\intfunk_k(\ve{p}(n))/\delta^n}$, $k\in\emenge{A}$, the SIR of user
    $k\in\emenge{A}$ evolves according to
    \begin{align*}
      \sir_k(n)&=
      \frac{\gamma_k(\ve{\pi}(n))_k}{(\intfun^{(a)}(\delta^n(\ve{\pi}(n),\ve{\lambda}))/\delta^n)_k}\geq
      \gamma_k
      ,\;n\geq N
    \end{align*}
    where the inequality follows from Proposition
    \ref{prop:ACL_ActiveStayActive}.  Hence, considering
    Lemma \ref{lem:convGeneralFun} yields (for all $n\geq N$)
    \begin{align}
      \label{eq:pi(n)boundedbelowByIa}
      \ve{\pi}(n)
      &\geq\intfun^{(a)}(\delta^n(\ve{\pi}(n),\ve{\lambda}))/\delta^n
      >\gintfun^{(a)}((\ve{\pi}(n),\ve{\lambda}))\,.
    \end{align}
    By Lemma \ref{lem:extendIntIsStandard},
    $\gintfun^{(a)}((\ve{\pi},\ve{\lambda}))$ is a standard interference
    function of $\ve{\pi}$. Thus, by \cite{Yates95a}, the function has a
    unique fixed point
    $\ve{\pi}^\ast=\gintfun^{(a)}((\ve{\pi}^\ast,\ve{\lambda}))>0$ such that
    $\ve{\pi}^\ast\leq\ve{\pi}$ for any
    $\ve{\pi}\geq\gintfun^{(a)}((\ve{\pi},\ve{\lambda}))$. So, from
    (\ref{eq:pi(n)boundedbelowByIa}) and $\tilde{A}$3, we have
    $\intfun^{(a)}(\delta^n(\ve{\pi}(n),\ve{\lambda}))/\delta^n>\gintfun^{(a)}((\ve{\pi}^\ast,
    \ve{\lambda}))$, which implies that
    \begin{equation}
      \label{eq:gammak<SIR<gammakSeq}
      \gamma_k\leq\sir_k(n)<
      \frac{\gamma_k(\ve{\pi}(n))_k}{(\gintfun^{(a)}((\ve{\pi}^\ast,\ve{\lambda})))_k},\; n\geq
      N,k\in\emenge{A}\,.
    \end{equation}
    Moreover, we have
    $\ve{\pi}(n)>\ve{\pi}^\ast=\gintfun^{(a)}((\ve{\pi}^\ast,\ve{\lambda}))>0$
    and $\ve{\pi}(n)<\ve{p}^{(a)}(0)$, where the last inequality is an
    immediate consequence of (\ref{eq:increasePowerActive_bounded}). Thus, the
    entries of $(\ve{\pi}(n),\ve{\lambda})$ are bounded and bounded away from
    zero. This together with Lemma \ref{lem:convGeneralFun} implies that, for each
    $k\in\emenge{A}$ and any $\epsilon>0$, there is (a sufficiently large)
    $M_k\geq N$ such that
    $1\leq(\intfunk_k(\delta^n(\ve{\pi}(n),\ve{\lambda}))/\delta^n)/
    \gintfunk_k((\ve{\pi}(n),\ve{\lambda}))<1+\epsilon$ for all $n\geq
    M_k$. So, by (\ref{eq:IterationPi(n+1)=Ia(pi(n))}), there is $M=\max_k
    M_k\geq N$ such that
    \begin{equation*}
      1\leq\max_{k\in\emenge{A}}\frac{(\ve{\pi}(n+1))_k}{
      \gintfunk_k((\ve{\pi}(n),\ve{\lambda}))}<1+\epsilon,\; n\geq M\,.
    \end{equation*}
    Now letting $\epsilon\to0$ ($n\to\infty$) shows that the sequence
    $\{\ve{\pi}(n)\}$ generated by (\ref{eq:IterationPi(n+1)=Ia(pi(n))})
    converges to
    $\ve{\pi}^\ast=\gintfun^{(a)}((\ve{\pi}^\ast,\ve{\lambda})$. Thus,
    from (\ref{eq:gammak<SIR<gammakSeq}), we obtain (as $n\to\infty$)
    \begin{equation*}
      \sir_k(n)
      =\frac{\gamma_k(\ve{\pi}(n))_k}{(\intfun^{(a)}(\delta^n(\ve{\pi}(n),\ve{\lambda}))
        /\delta^n)_k}\to
      \frac{\gamma_k\bar{p}_k}{\bar{p}_k}=\gamma_k,k\in\emenge{A}
    \end{equation*}
    where $\bar{p}_k=\lim_{n\to\infty}p_k(n)/\delta^n=(\ve{\pi}^\ast)_k
    =\gintfunk_k((\ve{\pi}^\ast,\ve{\lambda}))$.

\subsection{Proof of Proposition \ref{prop:ACL_Totallyadmissible}}
\label{app:proofTotallyAdmissible}

Let $\emenge{A}\neq\emptyset$ and $\emenge{B}$ be defined as in Proposition
\ref{prop:ACL_TotallyInadmissible}. The first part is proven by contradiction,
and hence assume that there exists 
$\ve{p}(0)$ for which
$\emenge{B}\neq\emptyset$. Thus, by Proposition
\ref{prop:ACL_TotallyInadmissible}, as $n\to\infty$,
  \begin{align}
  \label{eq:ACL_Totallyadmissible_proof1}
\begin{aligned}
  &\ve{p}^{(a)}(n)/\delta^n\to\ve{\pi}^\ast
  =\gintfun^{(a)}((\ve{\pi}^\ast,\ve{\lambda}))\\
  &\ve{p}^{(i)}(n)/\delta^n=\ve{\lambda}\leq\gintfun^{(i)}((\ve{\pi}^\ast,\ve{\lambda}))
  \end{aligned}
  &&
  \begin{aligned}
    &\ve{\pi}^\ast=\bar{\ve{p}}^{(a)}\\
    &\ve{\lambda}=\ve{p}^{(i)}(0)
  \end{aligned}
\end{align}
where $\gintfun^{(i)}$ is the (vector-valued)
interference function corresponding to the inactive users.
On the other hand, due to \ref{as:fullyadmissible}, we know from
\cite{Yates95a} that there exists $\ve{q}=(\ve{q}^{(a)},\ve{q}^{(i)})>0$ such
that $\ve{q}=\intfun(\ve{q})>\gintfun(\ve{q})$ where the last inequality is
due to \ref{lem:convGeneralFun}. Furthermore, using $\tilde{A}$2 we have 
$\mu\ve{q}>\gintfun(\mu\ve{q})$ for any $\mu>0$, and thus
\begin{align}
  \label{eq:ACL_Totallyadmissible_proof2}
\begin{aligned}  
  \mu\ve{q}^{(a)}>\gintfun^{(a)}((\mu\ve{q}^{(a)},\mu\ve{q}^{(i)}))
  \\
  \mu\ve{q}^{(i)}>\gintfun^{(i)}((\mu\ve{q}^{(a)},\mu\ve{q}^{(i)}))\,.
\end{aligned}  
\end{align}
Since $\ve{q}^{(i)}>0$ and $\ve{p}^{(i)}>0$, we can always choose the scaling
factor $\mu>0$ such that
$\ve{\lambda}\leq\mu\ve{q}^{(i)}$ with $\mu q_l^{(i)}=\lambda_l$ for some index $l$.
So, with \ref{as:NonOrthogonal}, (\ref{eq:ACL_Totallyadmissible_proof2}) and
our choice of $\mu$, we have
\begin{equation*}
  \gintfun^{(a)}((\mu\ve{q}^{(a)},\ve{\lambda}))
  \leq\gintfun^{(a)}((\mu\ve{q}^{(a)},\mu\ve{q}^{(i)}))
  <\mu\ve{q}^{(a)}\,.
\end{equation*}
As $\gintfun^{(a)}$ is a standard interference function (Lemma
\ref{lem:extendIntIsStandard}), $\ve{\pi}^\ast$ is a \emph{unique} fixed point
(given $\ve{\lambda}$) for which $\ve{\pi}^\ast\leq\ve{\pi}$ whenever
$\gintfun^{(a)}((\ve{\pi},\ve{\lambda}))\leq\ve{\pi}$. Thus, the above
inequality together with $\tilde{A}$3 and
(\ref{eq:ACL_Totallyadmissible_proof1}) implies that
\begin{align*}
  \ve{\pi}^\ast
  =\gintfun^{(a)}((\ve{\pi}^\ast,\ve{\lambda}))
  &\leq\gintfun^{(a)}((\mu\ve{q}^{(a)},\ve{\lambda}))
  <\mu\ve{q}^{(a)}\,.
\end{align*}
Hence, $\ve{\pi}^\ast<\mu\ve{q}^{(a)}$. Combining this with
(\ref{eq:ACL_Totallyadmissible_proof1}),
(\ref{eq:ACL_Totallyadmissible_proof2}) and $\tilde{A}$3 yields (for an index
$l$ such that $q_l^{(i)}=\lambda_l$)
\begin{align*}
  \lambda_l&\leq \gintfunk_l^{(i)}((\ve{\pi}^\ast,\ve{\lambda}))
  \leq\gintfunk_l^{(i)}((\mu\ve{q}^{(a)},\ve{\lambda}))
  \switchModes{}{\\&}\leq\gintfunk_l^{(i)}((\mu\ve{q}^{(a)},\mu\ve{q}^{(i)}))
  <\mu q_l^{(i)}=\lambda_l
\end{align*}
which is a contradiction. As a result, all users are admitted at some time
point $n_0$ and $\emenge{B}_{n}=\emptyset$ for all $n\geq n_0$. If all users
are admitted, the algorithm (\ref{eq:ActiveLinkProtection_ACL_Alg}) becomes a
pure fixed-point power control algorithm, and therefore, by
\ref{as:fullyadmissible} and \cite{Yates95a}, the power vector converges to
$\bar{\ve{p}}=\delta\intfun(\bar{\ve{p}})$.

\subsection{Proof of Proposition \ref{prop:ResultsUnderConstr}}
\label{app:proofResultsUnderConstr}

First we prove the following auxiliary result.
\begin{lemma}
Consider iteration (\ref{eq:AlgwithJn_PowerConstraints}). For each $k\in\logic$, 
there exists $n_0(k)$ such that for all $n \geq n_0(k)$ there holds:
\begin{equation}
	p_k(n+1)=\min\{\hat{p}_k, \delta\intfunk_k(\ve{p}(n))\}.
\label{eq:pcIterEquivToIntfunHat}	
\end{equation}
\end{lemma}
\begin{IEEEproof}
The lemma is proven by induction. Note that there exists $n=n_0(k)$ so that
\begin{align*}
p_k(n_0(k)+1)&=\delta\bigl(\aclfun(\ve{p}(n_0(k)),\hat{\ve{p}}/\delta)\bigr)_k
\switchModes{}{\\&}=\min\{\delta p_k(n_0(k)), \delta\intfunk_k(\ve{p}(n_0(k))),\hat{p}_k\}\\
&=\min\{\delta\intfunk_k(\ve{p}(n_0(k))),\hat{p}_k\}
\end{align*}
This can be immediately seen as otherwise there would hold $p_k(n+1)=\delta p_k(n)$ for all 
$n \geq 0$, and this is impossible due to power constraints. This proves the first step
of the induction. Now assume that \eqref{eq:pcIterEquivToIntfunHat}	is satisfied for some $n=j$.
We show that this implies that \eqref{eq:pcIterEquivToIntfunHat} holds for $n=j+1$ as well.
First consider the case (i): $p_k(j+1)=\hat{p}_k$. We get $p_k(j+2) = 
\min\{\delta \hat{p}_k, \delta \intfunk_k(\ve{p}(j)),\hat{p}_k\}=\min\{\delta\intfunk_k(\ve{p}(j+1)),\hat{p}_k\}$.
Now, assume (ii): $p_k(j+1)=\delta\intfunk_k(\ve{p}(j))$. In this case 
$p_k(j+2) = \min\{\delta^2\intfunk_k(\ve{p}(j)), \delta\intfunk_k(\ve{p}(j+1)),\hat{p}_k\}$.
Furthermore, $\ve{p}(j+1) \leq \delta \ve{p}(j)$, so considering properties A2 and A3 yields
$\delta\intfunk_k(\ve{p}(j+1)) \leq \delta\intfunk_k(\delta \ve{p}(j)) < \delta^2\intfunk_k(\ve{p}(j))$.
This in turn results in $p_k(j+2) = \min\{\delta\intfunk_k(\ve{p}(j+1)),\hat{p}_k\}$ also in the case (ii), 
which completes the second step of the induction and the proof of the lemma.
\end{IEEEproof}
The lemma implies that there exists $n_0$ such that for all $n \geq n_0$ the iteration
\eqref{eq:AlgwithJn_PowerConstraints} is equivalent to $\ve{p}(n+1)=\hat{\intfun}(\ve{p}(n))$ with
$\hat{\intfun}(\ve{p}):=\min\bigl\{\delta\intfun(\ve{p}), \hat{\ve{p}}\bigr\}$. The function 
$\hat{\intfun}(\ve{p})$ is a minimum of two standard interference functions and is therefore
a standard interference function itself, in the sense of Definition \ref{def:InterFunStandard}.
By \cite[Corollary 1]{Yates95a}, the iteration $\ve{p}(n+1)=\hat{\intfun}(\ve{p}(n))$ always
converges to a unique fixed point $\bar{\ve{p}}$. Finally, if \ref{as:fullyadmissible_PC} holds, 
then there exists $\ve{p}^\circ$ satisfying (\ref{eq:pdeltaMinimum_def}), 
which is the unique fixed point. This completes the proof.

\subsection{Proof of Proposition \ref{prop:pn<delta*p^ast}}
\label{app:proofpn<delta*p^ast}

Let $\ve{p}$ be any power vector satisfying
(\ref{eq:delta_valid_powervector}). Since \ref{as:fullyadmissible_PC} holds,
such a vector exists and
$\intfun(\delta\ve{p})<\delta\intfun(\ve{p})\leq\ve{p}\leq\hat{\ve{p}}$. Thus,
by continuity of $\intfun$, there is $\lambda>1$ such that
$\intfun(\lambda\delta\ve{p})\leq\hat{\ve{p}}$ with at least one equality. 

Now suppose that $\ve{p}(n)\leq\lambda\delta\ve{p}$ holds for some
$n\in\NNZ$. As
$p_k(n+1)=\min\{\hat{p}_k,\delta\intfun(\ve{p}(n))\},k\in\asetf_n$, and
$p_k(n+1)/\delta\leq p_k(n)<\intfunk_k(\ve{p}(n)),k\in\bsetf_n$, we then have
\begin{align*}
  \ve{p}(n+1)\leq\delta\intfunk(\ve{p}(n))\leq\delta\intfunk(\lambda\delta\ve{p})
  <\lambda\delta^2\intfunk(\ve{p})\leq\lambda\delta\ve{p}\,.
\end{align*}
Since this is true for any $n\in\NNZ$, we can conclude that if
(\ref{eq:pn<delta*p^ast}) is satisfied for some $m\in\NNZ$, then it holds for
all $n\geq m$.

Let $n\geq m$ be arbitrary. We are going to show that
$\asetf_n\subseteq\asetf_{n+1}$. By the above and $A2$, we have
\begin{enumerate}[(i)]
\item $\ve{p}(n)\leq\lambda\delta\ve{p}$ and
\item
  $\lambda\delta\ve{p}\geq\lambda\delta^2\intfun(\ve{p})\geq\delta\intfun(\lambda\delta\ve{p}),
  \lambda\geq 1$.
\end{enumerate}
Let $k\in\asetf_n$ be arbitrary and assume that $\hat{p}_k<\delta\gamma_k
I_k(\ve{p}(n)),k\in\asetf_n$. This does not impact the generality of the
analysis since otherwise the ALP property is provided. Due to (i), $A2$ and
$\ve{p}\leq\hat{\ve{p}}$, we have, for any $k\in\asetf_n$,
\begin{align*}
  \sir_k(\ve{p}&(n+1))\geq\frac{\gamma_k I_k(\lambda\delta\ve{p})}
  {I_k(\min\{\delta\ve{p}(n),\delta\intfun(\ve{p}(n))\})}
  \switchModes{}{\\&}\geq\frac{\gamma_k I_k(\lambda\delta\ve{p})}
  {I_k(\min\{\lambda\delta^2\ve{p},\delta\intfun(\lambda\delta\ve{p})\})}
  \geq\frac{\gamma_k I_k(\lambda\delta\ve{p})}
  {I_k(\delta\intfun(\lambda\delta\ve{p}))}\,.
\end{align*}
Thus, by (ii) and $A3$, we have $I_k(\lambda\delta\ve{p})\geq
I_k(\delta\intfun(\lambda\delta\ve{p})),k\in\asetf_n$, so that
$\sir_k(\ve{p}(n+1))>\gamma_k$ or, equivalently, $k\in\asetf_{n+1}$. Since
this is true for any $k\in\asetf_n$, we obtain
$\asetf_n\subseteq\asetf_{n+1}$. By the preservation of
(\ref{eq:pn<delta*p^ast}), we can finally conclude that
$\asetf_{n}\subseteq\asetf_{n+1}$ for all $n\geq m$.

\subsection{Proof of Proposition \ref{prop:pn<deltaIn_protection}}
\label{app:pn<deltaIn_protection_proof}

We first prove the following lemma.
\begin{lemma}
\label{lem:p_ast_isMinimum}
Suppose that \ref{as:fullyadmissible_PC} holds and
$\ve{p}\leq\delta\intfun(\ve{p})$ for some $\ve{p}>0$. Then,
$\ve{p}\leq\ve{p}^\circ$, where $\ve{p}^\circ$ is defined by
(\ref{eq:pdeltaMinimum_def}).
\end{lemma}
\begin{IEEEproof}
  The proof is by contradiction. Thus, assume that $p_k>p_k^\circ$ for some
  $k\in\logic$. But, as $\ve{p}$ and $\ve{p}^\circ$ are both positive vectors,
  this implies that there exists $\mu>1$ such that $\mu\ve{p}^\circ\geq\ve{p}$
  and $\mu p^\circ_l=p_l$ for some $l\in\logic$. Hence, by
  (\ref{eq:pdeltaMinimum_def}), $A2$ and $A3$, we have $\mu
  p^\circ_l=\mu\intfunk_l(\ve{p}^\circ)>\intfunk_l(\mu\ve{p}^\circ)\geq\intfunk_l(\ve{p})\geq
  p_l$, which contradicts $\mu p_l^\circ=p_l$.
\end{IEEEproof}

Now, since
\ref{as:fullyadmissible_PC} is assumed to hold, there exists $\ve{p}$ with
(\ref{eq:delta_valid_powervector}).  By $A2$, we further have
$\intfun(\lambda\delta\ve{p}) \leq\hat{\ve{p}}$ for some $\lambda\geq 1$ (and,
in fact, $\lambda>1$). Let $n=m\in\NNZ$ be any time point for which
(\ref{eq:pn<lambda_deltaIn/lambda}) is fulfilled, and let $k\in\asetf_{n}$ be
arbitrary. We can assume $\hat{p}_k<\delta\intfunk_k(\ve{p}(n))$, so it
follows from (\ref{eq:SIR(n+1)withequality}) and $A3$ that
\begin{align}
\label{eq:pn<deltaIn_protection_proof_1}
\sir_k(\ve{p}(n+1))&\geq\frac{\hat{p}_k}{I_k(\ve{p}(n+1))} \geq
\gamma_k\frac{I_k(\lambda\delta\ve{p})}{I_k(\ve{p}(n+1))}
\end{align}
Now (\ref{eq:pn<lambda_deltaIn/lambda}) together with Lemma
\ref{lem:p_ast_isMinimum} and \cite[Lemma 1]{Yates95a} implies that
\begin{equation}
\label{eq:pn<deltaIn_protection_proof_2}
\frac{\ve{p}(n)}{\lambda\delta}\leq\ve{p}^\circ\leq\ve{p}
\end{equation}
where $\ve{p}^\circ$ is defined by (\ref{eq:pdeltaMinimum_def}).  Proposition
\ref{prop:pn<delta*p^ast} implies that $\ve{p}(n+1)\leq\lambda\delta\ve{p}$
for any $\delta$-valid power vector $\ve{p}$. Thus, by $A3$, we have
$\intfun(\ve{p}(n+1))\leq\intfun(\lambda\delta\ve{p})$, from which and
(\ref{eq:pn<deltaIn_protection_proof_1}) one obtains
$\sir_k(\ve{p}(n+1))\geq\gamma_k$. Thus, $k\in\asetf_{n+1}$ and
$\asetf_n\subseteq\asetf_{n+1}$ as $k\in\asetf_n$ is arbitrary.

Finally, Proposition \ref{prop:pn<delta*p^ast} shows that
(\ref{eq:pn<delta*p^ast}) (or, equivalently,
(\ref{eq:pn<deltaIn_protection_proof_2}) with $n=m$) is preserved for all
$n\geq m$. Thus, $\asetf_n\subseteq\asetf_{n+1}$ for all $n\geq m$, which
completes the proof.

\subsection{Proof of Proposition \ref{prop:pn<delta2In_protection_always}}
\label{app:pn<delta2In_protection_always_proof}

First consider the following simple lemma.

\begin{lemma}
\label{lem:SIRn+1_bounded_SIRn}
Let $\beta\geq 1$ be arbitrary. If (\ref{eq:pn<delta2In/lambda_always}) holds
for some $m\in\NNZ$, then $\ve{p}(n)\leq\beta\delta\intfun\bigl(\ve{p}(n))$
for all $n\geq m$.
\end{lemma}
\begin{IEEEproof}
  Let $n\in\NNZ$ be any natural number for which
  (\ref{eq:pn<delta2In/lambda_always}) holds with $m=n$. We are going to show
  that (\ref{eq:pn<delta2In/lambda_always}) is satisfied for $m=n+1$. If
  (\ref{eq:pn<delta2In/lambda_always}) holds with $m=n$, then, by
  (\ref{eq:AlgwithJn_PowerConstraints}) and $p_k(n)\leq\hat{p}_k$, one obtains
  \begin{equation}
    \label{eq:pn<delta2In/lambda_always_proof}
    \begin{split}
    \ve{p}(n+1)&=\min\bigl\{\delta
    \ve{p}(n),\delta\intfun(\ve{p}(n)),\hat{\ve{p}}\bigr\}
    \switchModes{}{\\&}\geq\min\bigl\{
    \ve{p}(n),\delta\intfun(\ve{p}(n)),\hat{\ve{p}}\bigr\}\\
    &\geq\min\bigl\{
    \ve{p}(n),\delta\intfun(\ve{p}(n))\bigr\}\geq\ve{p}(n)/\beta\,.
    \end{split}
  \end{equation}
  On the other hand, it follows from (\ref{eq:AlgwithJn_PowerConstraints})
  that $p_k(n+1)\leq\delta\intfunk_k(\ve{p}(n))$ for any $k\in\asetf_n$. If
  $k\in\bsetf_n$, then $p_k(n+1)=\delta
  p_k(n)\leq\delta\intfunk_k(\ve{p}(n))$, where the last step follows from the
  fact that $p_k(n)\leq\intfunk_k(\ve{p}(n))$ for each $k\in\bsetf_n$. Thus,
  $\ve{p}(n+1)\leq\delta\intfun(\ve{p}(n))$. When combined with
  (\ref{eq:pn<delta2In/lambda_always_proof}) and $A2$, this yields
  $\ve{p}(n+1)\leq\delta\intfun(\beta\ve{p}(n+1))<\beta\delta\intfun(\ve{p}(n+1))$. 
\end{IEEEproof}

Now we use the lemma to prove the proposition. To this end, let $m\in\NNZ$ be
any time point for which (\ref{eq:pn<delta2In/lambda_always}) holds and define
\begin{equation*}
  \beta_{\max}=\min_{k\in\logic}\frac{\delta\lambda\intfunk_k(\ve{p}(m)/\delta\lambda)}{\intfunk_k(\ve{p}(m))}
\end{equation*}
where $\lambda>1$ is defined in Proposition
\ref{prop:pn<deltaIn_protection}. By (\ref{eq:pn<delta2In/lambda_always}), one
has $\ve{p}(m)\leq
\beta\delta\intfun(\ve{p}(m))\leq\lambda\delta^2\intfun(\ve{p}(m)/\delta\lambda)$
for all $\beta\in[1,\beta_{\max}]$.  Proposition
\ref{prop:pn<deltaIn_protection} ensures the protection at time $m$. Moreover,
by Lemma \ref{lem:SIRn+1_bounded_SIRn}, the condition
(\ref{eq:pn<delta2In/lambda_always}) is preserved for all $n\geq m$ so that
$\aset_{n}\subseteq\asetf_{n+1}$ for all $n\geq m$. Finally, since $1\leq a<b$
implies
\begin{align*}
  a\intfun(\ve{p}/a) =a\frac{b}{a}\frac{a}{b} \intfun(\ve{p}/a)
  \overset{A2}{<}b \intfun\bigl(\frac{a}{b}\frac{\ve{p}}{a}\bigr)
  =b\intfun(\ve{p}/b)
\end{align*}
for any fixed $\ve{p}$, the function
$\RP\to\RP:x\mapsto x\intfun(\ve{p}(m)/x)$ is strictly increasing and, by
Proposition \ref{prop:InterFunContinuous}, continuous in every
component. Thus, we have $\beta_{\max}>1$ as $\delta\lambda>1$, which completes the
proof.

\bibliographystyle{IEEEbib}
\bibliography{StaKalBam2008JournalAdmissionControl.bbl}

\end{document}